\pdfoutput=1

\documentclass[reprint,
superscriptaddress,
 amsmath,amssymb,
 aps,
 prx,
]{revtex4-2}

\usepackage{graphicx}%
\usepackage{dcolumn}%
\usepackage{verbatim}
\usepackage{xcolor}
\usepackage{bm}%
\usepackage{braket}
\usepackage{hyperref}%
\usepackage{booktabs}
\usepackage{array}%
\makeatletter
\AtBeginDocument{%
  \let\AHT@rv@mkpream\@mkpream
  \def\@mkpream#1{%
    \let\@endpbox\relax
    \let\do@row@strut\relax
    \let\ar@align@mcell\relax
    \AHT@rv@mkpream{#1}}%
}
\makeatother
\hypersetup{
    colorlinks,
    linkcolor={green!80!black},
    citecolor={blue!100!black},
    urlcolor={blue!100!black}
}
\usepackage[labelformat=simple]{subcaption} %

\usepackage{siunitx}
\usepackage{quantikz}

\tikzset{
  quantum gate/.append style={rounded corners}
}

\newcommand{\intf}[4]{\int_{#1}^{#2} \! #3 \, \mathrm{d}#4}
\newcommand{\sotimes}{\!\otimes\!}

\newcommand{\ketbra}[2]{\ket{#1}\!\bra{#2}}

\makeatletter
\def\imod#1{\allowbreak\mkern10mu({\operator@font mod}\,\,#1)}
\makeatother

\DeclareCaptionLabelFormat{nocaption}{}
\usepackage{ragged2e}
\DeclareCaptionJustification{justified}{\justifying}

\newif\ifmarkrevisions

\markrevisionsfalse
\newcommand{\revision}[1]{
\ifmarkrevisions
\textcolor{blue}{#1}
\else
{#1}
\fi}

\begin{document}

\preprint{APS/123-QED}

\title{Hybrid analog-digital simulation of the Abelian Higgs model with transmon qutrits}%

\author{Muhammad Asaduzzaman}\thanks{These authors contributed equally to this work.}
\affiliation{Department of Physics and Astronomy, North Carolina State University,  Raleigh, NC 27607, USA}
\affiliation{Department of Physics and Astronomy, The University of Iowa, Iowa City, IA 52242, USA}
\author{Rayleigh W. Parker}\thanks{These authors contributed equally to this work.}
\affiliation{Department of Physics and Astronomy, University of Rochester, Rochester, New York 14627, USA}
\affiliation{University of Rochester Center for Coherence and Quantum Science, Rochester, New York 14627, USA}
\author{Noah Goss}
\affiliation{Department of Physics, University of California, Berkeley, Berkeley, CA 94720, USA.}
\affiliation{Computational Research Division, Lawrence Berkeley National Laboratory, Berkeley, CA 94720, USA.}
\author{Ahmed I. Mohamed}
\affiliation{Department of Physics and Astronomy, University of Rochester, Rochester, New York 14627, USA}
\affiliation{University of Rochester Center for Coherence and Quantum Science, Rochester, New York 14627, USA}
\author{Max Neiderbach}
\affiliation{Department of Physics and Astronomy, University of Rochester, Rochester, New York 14627, USA}
\affiliation{Department of Physics, University of Maryland,  4150 Campus Dr, College Park, MD 20742, USA}
\author{Zane Ozzello}
\affiliation{Department of Physics and Astronomy, The University of Iowa, Iowa City, IA 52242, USA}
\author{Ravi K. Naik}
\affiliation{Computational Research Division, Lawrence Berkeley National Laboratory, Berkeley, CA 94720, USA.}
\author{Alexander F. Kemper}
\affiliation{Department of Physics and Astronomy, North Carolina State University, Raleigh, NC 27607, USA}
\author{Irfan Siddiqi}
\affiliation{Department of Physics, University of California, Berkeley, Berkeley, CA 94720, USA.}
\affiliation{Computational Research Division, Lawrence Berkeley National Laboratory, Berkeley, CA 94720, USA.}
\author{Yannick Meurice}
\affiliation{Department of Physics and Astronomy, The University of Iowa, Iowa City, IA 52242, USA}
\author{Machiel S. Blok}\thanks{Contact author: machielblok@rochester.edu}
\affiliation{Department of Physics and Astronomy, University of Rochester, Rochester, New York 14627, USA}
\affiliation{University of Rochester Center for Coherence and Quantum Science, Rochester, New York 14627, USA}

\date{\today}%

\begin{abstract}
To investigate gauge theories with near-term quantum computers warrants exploration of nontraditional quantum simulators to find resource-efficient simulation protocols and ultimately access exotic features of different field theories, including unexplored regimes of the QCD phase diagram. In this work, using superconducting transmon qutrit processors, we formulate and implement a pulse-based, three-level, hybrid analog-digital simulation protocol of the (1+1) dimensional Abelian Higgs model (AHM) on two sites. Alongside this approach, we experimentally realize a gate-based implementation of the same model. Using the natural mapping of the three-level truncation of the transmon Hilbert space to the spin-1 truncated AHM, we observe real time dynamics of AHM field observables, which are analogous to electric field operators, with both protocols. For the analog-digital protocol, we engineer a Floquet simulation with a combination of local analog drives, driven modification of the natural interaction Hamiltonian of the two transmons, and dynamical decoupling pulses. For the digital protocol, we use a state-of-the-art qutrit processor to implement a Trotterized simulation of the model incorporating  advanced error mitigation techniques.
We further discuss the scalability of the two approaches, and their potential to be extended to the simulation of other model Hamiltonians. Our experiments demonstrate a viable platform for future studies of spin-1 and SU(3) based gauge theory models on current and near-term transmon qutrit processors.
\end{abstract}
\maketitle
\section{Introduction}
The prospect of simulating quantum field theories with quantum simulators has attracted considerable interest in recent years. While classical numerical methods based on importance sampling have successfully established lattice quantum chromodynamics (QCD) with imaginary time as a reliable nonperturbative formulation of strong interactions, they face significant challenges in addressing the \emph{sign problem} that appears in the presence of large chemical potential \cite{Troyer:2004ge,Gattringer:2016kco} and the difficulty of observing real-time non-equilibrium dynamics \cite{Zohar:2021nyc}.
In light of these difficulties, there have been several efforts to simulate lattice gauge theories with various quantum computing platforms including superconducting circuits, trapped ions, and Rydberg atom simulators in (1+1) dimensions \cite{Charles:2023zbl,Bazavov:2015kka,Martinez:2016yna,De:2024smi,Luo:2025qlg,Jha:2024jan} and (2+1) dimensions \cite{Lamm:2019bik,Gonzalez-Cuadra:2024xul,Yamamoto:2020eqi,Ding:2021ysm,Maiti:2024jwk,Mueller:2024mmk,Sukeno:2022pmx,Homeier:2022mkg, meth_simulating_2025, cochran_visualizing_2025}. These studies mark essential intermediate steps towards simulating QCD in (3+1) dimensions, and they address diverse physical phenomena including phase transitions \cite{Rosanowski:2025nck}, scattering of particles \cite{Jha:2024jan}, real-time confining dynamics \cite{De:2024smi, cochran_visualizing_2025}, quench dynamics \cite{Xiang:2025qhq}, the role of entanglement in thermalization \cite{Zhou:2021kdl}, and quantum chaos \cite{Mueller:2024mmk}.

So far, most studies have been performed on platforms composed of quantum two-level systems, or qubits. Platforms made of \emph{qudits}, or quantum $d$-level systems, however, have the potential to reduce the gate depths and non-local quantum operations required for quantum simulations of the naturally $d$-level systems that are common in lattice field theory models \cite{jiang_non-abelian_2025,Kurkcuoglu:2021dnw,gustafson_noise_2022,illa_qu8its_2024}.
And yet, despite recent successes in the development of qudit processors with superconducting circuits \cite{Blok:2020may, roy_two-qutrit_2023}, trapped ions \cite{ringbauer_universal_2022}, and photonic systems \cite{chi_programmable_2022}, and despite a number of theoretical proposals to study field theories with qudits \cite{Ciavarella:2021nmj,
gustafson_prospects_2021, Meurice:2021pvj,gonzalez-cuadra_hardware_2022,zache_fermion-qudit_2023,illa_quantum_2023,popov_variational_2024,illa_qu8its_2024, calajo_digital_2024, calliari_quantum_2025,jiang_non-abelian_2025}, \revision{to our knowledge, there have been only two experimental realizations of qudit-based field theory simulations in the literature so far, which were demonstrated on trapped-ion quantum hardware \cite{meth_simulating_2025, john_non-abelian_2026}}.

\begin{figure*}[!htb]
    \captionsetup[subfigure]{labelformat=nocaption}
    \centering
    \includegraphics[width=0.9\linewidth]{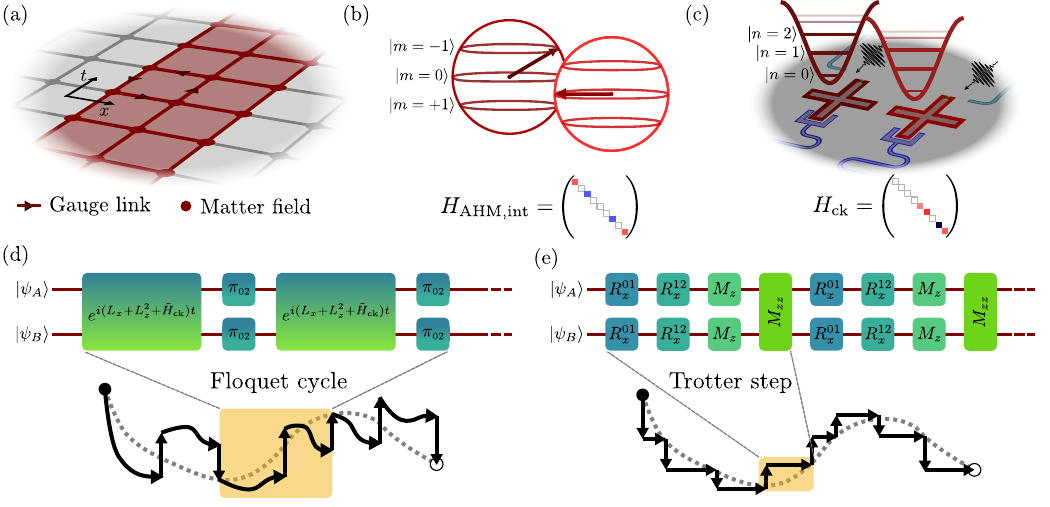}
    \begin{subfigure}{0\linewidth}
    \caption{}\label{fig:abelian_higgs_lattice_cartoon}
    \end{subfigure}%
    \begin{subfigure}{0\linewidth}
    \caption{}\label{fig:abelian_higgs_spin_cartoon}
    \end{subfigure}%
    \begin{subfigure}{0\linewidth}
    \caption{}\label{fig:transmon_cartoon}
    \end{subfigure}%
    \caption{ (a) Lattice field theory representation of the Abelian Higgs model (AHM) in (1+1) dimensions. Vertices represent matter fields and edges represent gauge links in the discretized spacetime. In this work we consider a time-slice that is two plaquettes wide. (b) In the appropriate limits and approximations (see Ref.~\cite{Bazavov:2015kka}), the (1+1)D AHM can be mapped to a spin-1 chain, where the spin states correspond to the plaquette quantum numbers. We illustrate the two spin-1 particles we simulate in this experiment along with the shape of their interaction Hamiltonian $H_\mathrm{AHM,int}\propto L_z\sotimes L_z$. (c) Our experimental platform consists of a pair of transmon qutrits, whose native cross-Kerr interaction $H_\mathrm{ck}$ is diagonal but differs greatly in form from the target interaction. By applying microwave drives to the two transmons, we may approximate evolution under the model Hamiltonian in two distinct ways. (d) An analog-digital hybrid approach to engineering the evolution of the target Hamiltonian. Here we alternate between evolution under a Hamiltonian engineered with analog drives and approximately instantaneous local $\pi_{02}$ gates, which transpose the states $\ket{0}$ and $\ket{2}$ on each transmon individually. (e) The fully digital approach. Here a sequence of digital gates drawn from a universal set is repeated to effectively Trotterize the time evolution. Definitions of the gates are given in Sec.~\ref{sec:gates}. For both protocols, a sketch of the trajectory of the state is shown. In the analog-digital simulation, the evolution can deviate substantially from the ideal trajectory (dashed line) but returns after a full Floquet cycle up to the error in the Magnus expansion. In the digital simulation, the evolution is closer to the exact solution throughout the evolution, and approximates it up to Trotter error at the end of each Trotter step.}
    \label{fig:figure1}
\end{figure*}

In this work, we consider a particular lattice field theory model---the spin-1 truncated Abelian Higgs model (AHM) \cite{Peskin:1995ev,Fradkin:1978dv,einhorn1979phase,callaway1982abelian}  in two spacetime dimensions---and perform qudit-based quantum simulations of the model in superconducting circuit platforms with two separate but complementary approaches (Fig.~\ref{fig:figure1}). In the first, we leverage Hamiltonian engineering techniques to perform an analog-digital hybrid simulation, while in the second, we perform a fully digital simulation, taking advantage of state-of-the-art error mitigation methods, finding a nearly fivefold reduction in entangling-gate depth compared to a qubit-based digital simulation.
The AHM is a useful testbed for proof-of-principle simulators, not least because of its shared features with QCD, including confinement \cite{polyakov1977quark,Schaposnik:1978wq}, and string breaking \cite{Fradkin:1978dv}, making it a valuable step towards quantum simulation of QCD. \revision{Here we implement quantum simulations of the model on two lattice sites to demonstrate the fundamental building blocks as a first step towards observing many-body phenomena in the Abelian Higgs model.}
Owing to the spin-1 truncation of the model, each site can be naturally mapped to a qutrit ($d=3$), which in both simulation methods we encode in a truncated subspace spanned by the lowest three energy eigenstates of a transmon. Transmons are superconducting anharmonic oscillators whose nonlinearity enables high fidelity universal control over a portion of their Hilbert space \cite{Koch:2007hay}.
Although they were initially engineered to implement qubits, they are seeing increasing use in higher-dimensional quantum computation and simulation \cite{bianchetti_control_2010, Blok:2020may, roy_two-qutrit_2023, cho_direct_2024, kumaran_transmon_2025, ticea_observation_2026}.

In our first simulation approach, we engineer the Hamiltonian of a two-transmon system to approximate that of the target model with a hybrid analog-digital protocol, producing an effective Floquet Hamiltonian by alternating between periods of analog evolution and relatively short sequences of digital gates \cite{hayes_programmable_2014, arrazola_digital-analog_2016, choi_dynamical_2017}. Similar methods have been used for previous quantum simulations in a wide variety of platforms \cite{langford_experimentally_2017,babukhin_hybrid_2020, andersen_thermalization_2025,Choi2020, Zhou2024,morong_engineering_2023,scholl_microwave_2022}. These experiments have demonstrated analog-digital hybrid methods to be a powerful tool for quantum simulation, as they enable the reduction or elimination of relatively low-fidelity two-qutrit gates, while retaining some of the flexibility provided by fine-tuned digital control.

In our second approach, we perform a digital gate-based simulation of our model on a pair of transmons in a universal qutrit processor \cite{Morvan:2020rrb,Goss:2023frd}. Compared to a qubit-based gate decomposition of the model, we observe a four-fold reduction in resource requirements when decomposing the desired entangling operation with qutrit-based quantum gates. We incorporate state-of-the-art error mitigation protocols including randomized compilation and purification methods to compute observables that in future studies could identify phase transitions in large lattices. Our investigation demonstrates that resource requirements for digital error mitigation with twirling do not increase significantly when qubits are replaced with qutrits as the computational unit, in agreement with previous studies \cite{Goss:2023frd}.

This paper is structured as follows. In Sec.~\ref{sec:model}, we review an experimentally accessible lattice Hamiltonian formulation of the Abelian Higgs model.  In Sec~\ref{sec:hybrid}, we introduce the analog simulation approach for a single-site AHM and an analog-digital hybrid approach for the multi-site AHM Hamiltonian. We discuss the practical considerations for designing and implementing this protocol and present the results obtained in experiment. Next, in Sec.~\ref{sec:gates} we decompose the unitary operation generated by the AHM Hamiltonian into the native gate set of our processor. We present the results obtained from the device and discuss the implications of different error mitigation strategies in our digital quantum simulation. Finally, we discuss the trade-offs and scalability of the two approaches in Sec~\ref{sec:interpretation} and present our conclusions in Sec~\ref{sec:conclusions}.

\vskip5pt

\section{Abelian-Higgs model}\label{sec:model}
The Abelian Higgs model is a toy model of scalar quantum electrodynamics that describes the interaction of a photon field with spin zero particles.
We begin with a brief description of the model and its approximate representation as a spin-1 Hamiltonian on the lattice.
The action for the model is described by a $U(1)$ gauge field and a complex matter field $\Phi$ \cite{einhorn1979phase,Banks:1979fi,Meurice:2021pvj}

\begin{equation}
S=\int \mathrm{d}^2 x \frac{1}{4} \mathcal{F}_{\mu \nu} \mathcal{F}^{\mu \nu} +\left|D_\mu \Phi\right|^2+\frac{\mu^2}{2}\Phi^\dagger \Phi-\frac{\lambda}{4} (\Phi^\dagger \Phi)^2, \label{eqn_AH}
\end{equation}
where $\mathcal{F}_{\mu\nu} =\partial_\mu A_\nu -\partial_\nu A_\mu$ is the electromagnetic tensor and  $D_\mu =\partial_\mu +iqA_\nu$ is the covariant derivative with matter charge $q$. The model is invariant under the $U(1)$ gauge transformation $\Phi \to e^{i\alpha(x)}\Phi,\,\, A_\mu \to A_\mu -\frac{1}{q}  \partial_\mu \alpha(x)$.
We work in the limit where the Higgs mode is decoupled and the complex scalar field is reduced to a constant multiplied by a complex phase (the Nambu-Goldstone mode). In (1+1) dimensions, this model has no deconfining phase; in higher dimensions, however, it does exhibit a similar confining feature of a charge and anti-charge pair \cite{Fradkin:1978dv,einhorn1979phase}. Confinement is marked by a linear growth in potential due to static charge separation below a critical distance scale and string breaking phenomena is observed at large distances signifying charge screening \cite{polyakov2018gauge}. In real-time quantum simulators, both the confining feature and string breaking phenomena can be observed from the local charge density \cite{Halimeh:2025vvp}. The shape of the light cone can identify confining features in the model and string breaking can be identified from the integrated local charge density by counting the number of charges present.

The action is translated into a discretized form by placing the matter fields on the vertices and gauge fields on the links of a 2D lattice---as illustrated in Fig.~\ref{fig:abelian_higgs_lattice_cartoon}---which can also be used for classical numerical simulation of the model using the path integral formulation. Taking the time continuum limit of the lattice action following the steps outlined in Ref.~\cite{Bazavov:2015kka}, we obtain the lattice Hamiltonian on $N_s$ sites, assuming open boundary conditions,
\begin{equation}
\begin{aligned}
    H_{\mathrm{AHM}} &=\frac{\kappa}{2} \sum_{k=1}^{N_s}\left(L^{(k)}_z\right)^2 -\chi \sum_{k=1}^{N_s} F^{(k)}_x\\
     &+\frac{\beta}{2} \sum_{k=1}^{N_s-1}\left(L^{(k+1)}_z-L^{(k)}_z\right)^2.\label{eqn_latticeModel}
\end{aligned}
\end{equation}
We define $\hbar$ = 1 throughout this work. The $L_z$ operator appearing in the Hamiltonian formulation is the canonical conjugate of the gauge field and is analogous to electric field operator in quantum electrodynamics, whereas charge corresponds to the conjugate of the phase operator of the Higgs field. In this (1+1) dimensional model, we do not need to account for the matter fields explicitly, as the charge can be computed as the difference of the electric field operators ($L_z^{k+1}-L_z^k$), due to Gauss's Law.

\revision{As a consequence, the local operators $L_z^{(k)}$, where $k$ indexes the site that each operator acts on, provide a convenient basis in which to analyze the behavior of the model, and will be used as our primary observables in this work. In future, larger scale studies, monitoring the local charge through these observables could enable direct investigations of string-breaking phenomena.}

\revision{The terms in the Hamiltonian can be interpreted, respectively, as the local energy of the electric field; an injection of energy that drives local evolution of the field; and the cost in energy due to creation of charges. The operators $L_z^{(k)}$ are identical to the angular momentum operator defined by
\begin{equation}
        L_z|m\rangle = m|m\rangle.
\end{equation}
We may map spin $L_z$ eigenstates to transmon eigenstates with
 \begin{equation}
 \begin{aligned}
    &\ket{m=1} \to \ket{n=0},\quad
    \ket{m=0} \to \ket{n=1},  \\
    &\ket{m=-1}\to \ket{n=2}.\label{eq:qutrit_mapping}
\end{aligned}
\end{equation}
Measuring $L_z$ expectation values then corresponds to a measurement of local qutrit state populations.}

The operators $F^{(k)}_x$ are defined by the following equations:
\begin{equation}
        F_{ \pm}|m\rangle =|m \pm 1\rangle,  \hspace{2pt}
        F_x =\frac{1}{2}\left(F_{+}+F_{-}\right). \label{eqn_fund_op}
\end{equation}
To capture the model exactly, the operators $L_z^{(k)}$ and $F_x^{(k)}$
are required to act on infinite levels: $m \in \{ 0, \pm 1, \pm 2, \cdots, \pm \infty \}$, but in practice, some cutoff $|m|\leq l$ is required. In this work we take $l=1$, and in this case we may identify $\sqrt{2}F_x$ with the spin-1 $L_x$ operator.

This spin-1 approximation allows for the development of new tools for simulating various gauge theory models with a more efficient encoding than is available on qubit platforms. \revision{As is discussed further in Sec \ref{sec:gates} and Appendix \ref{app_qubit_resource_estimate}, performing a digital quantum simulation of this truncated Abelian Higgs model on qubit-based hardware would require   twice the number of qubits as qutrits, and would involve greatly increased gate depths because many single-site operations would effectively become two-qubit operations. In the following two sections, we demonstrate two methods to more directly and efficiently implement the model with transmon qutrits.}

\section{Analog-Digital Simulation}\label{sec:hybrid}
We first discuss the analog-digital simulation of the Abelian Higgs model on two sites ($N_s=2$). Experiments in this section were performed on a Rochester-based transmon device with properties given in Table~\ref{tab:analog_coherence}. The implementation can be broken into two parts: (1) driving single-transmon dynamics to emulate local on-site terms in the model, and (2) engineering the interaction between the transmons to effectively match the $L_z \otimes L_z$ interaction term between the sites in the model. To reduce Trotterization errors and to maximize simulation time, we aim to perform these operations simultaneously, which introduces an additional challenge: designing the drives for the single-site dynamics and the interaction-engineering in such a way that each does not interfere with the other's operation.

\subsection{Single-qutrit implementation}\label{sec:single_qutrit_analog}
We begin by discussing the analog drives that produce the desired single-transmon Hamiltonian at each site. In the frame of the lab, the Hamiltonian of transmon $k$ truncated to three levels with a time-dependent drive $V_k(t)$ may be written \cite{Koch:2007hay}
\begin{equation}
\begin{aligned}
    H_k^\mathrm{LF}(t) &= H^0_k + H_k^\mathrm{drive}(t)  \\
    &= \omega^k_{01} \ketbra{1}{1} +(\omega^k_{01} +\omega^k_{12}) \ketbra{2}{2} \\
    &+ V_k(t) \left(i\ketbra{1}{0} + i\lambda_k \ketbra{2}{1}  + \mathrm{h.c.}\right),
    \label{eq:single_transmon_lab_frame_1}
\end{aligned}
\end{equation}
where $\omega^k_{ij}$ is defined as the transition frequency between transmon eigenstates $\ket{i}$ and $\ket{j}$, %
and $\lambda_k\approx\sqrt{2}$ is a dimensionless factor which characterizes the ratio between the charge couplings to the $\ket{1}\leftrightarrow\ket{2}$ transition and the $\ket{0}\leftrightarrow\ket{1}$ transition. For the remainder of this subsection, we will suppress the transmon index $k$ on terms in the Hamiltonian.

We drive both transitions simultaneously, taking $V(t) = \sum_{i=1}^{2}\Omega_i(t)\cos(\omega_{i}^\mathrm{d}t + \phi_i)$ where $\Omega_i$ are the drive envelopes, $\omega_{i}^\mathrm{d}$ are the drive frequencies, and $\phi_i$ are the phases on each drive term. We further define detunings $\Delta_i = \omega_{i}^\mathrm{d}-\omega_{i-1,i}$ and assume $|\Delta_i|+\Omega_i \ll |\omega_{12}-\omega_{01}|$, so that each drive term is close to resonance with only one transition. For both transmons in the analog-digital experiment, the anharmonicity was roughly $\left(\omega_{12}-\omega_{01}\right)/2\pi\sim \qty{-240}{\MHz}$.

Extracting the on-site terms from the model Hamiltonian in Eq.~\eqref{eqn_latticeModel}, we find
\begin{equation}
		H_{\mathrm{AHM},k} =  \frac{\kappa+\beta}{2} L_z^2-\frac{\chi}{\sqrt{2}} L_x,\label{eq:single_qutrit_model}
\end{equation}
where site $k$ is taken to be at the boundary of the chain for both sites of the two-site model. To find the drives required to simulate $H_{\mathrm{AHM},k}$, we first perform a unitary rotating frame transformation to take $H_k^\mathrm{LF}$ into the frame of the drive frequencies: $H_k^\mathrm{DF} = U_{\mathrm{D}}^\dagger H_{k}^\mathrm{LF} U_{\mathrm{D}}-i U_{\mathrm{D}}^\dagger \partial_t U_{\mathrm{D}}$, where $U_{\mathrm{D}}(t)=\mathrm{diag}\left(1,e^{-i\omega_{1}^\mathrm{d} t}, e^{-i (\omega_{1}^\mathrm{d} + \omega_{2}^\mathrm{d}) t}\right)$ \cite{li_decoherence_2011,krantz_quantum_2019}. Since this frame transformation is diagonal, the population dynamics we observe in the lab frame will be identical to the population dynamics in the rotating frame. After applying the rotating wave approximation (RWA) under the assumptions that the drive strengths are weak compared to the drive frequencies and compared to the anharmonicity of the transmon, $\Omega_1,\Omega_2 \ll \left|\omega_{01}-\omega_{12}\right|<\omega_{1}^\mathrm{d},\omega_{2}^\mathrm{d}$, we obtain,
\begin{equation}
\begin{aligned}
		&H_k^\mathrm{DF}(t) \approx -\Delta_{1} \ketbra{1}{1} - (\Delta_{1}+\Delta_{2})\ketbra{2}{2}\\
        &+\frac{i}{2}\Omega_1(t)e^{-i\phi_1} \ketbra{1}{0}
        + \frac{i\lambda}{2}\Omega_2(t)e^{-i \phi_2} \ketbra{2}{1}
			 + \mathrm{h.c}.
		\label{eq:H_rwa}
\end{aligned}
\end{equation}
More details of the terms dropped under the RWA can be found in Appendix~\ref{app:rotating_frame}.

\begin{figure}[!htb]
\captionsetup[subfigure]{labelformat=nocaption}
\includegraphics[width=0.9\linewidth]{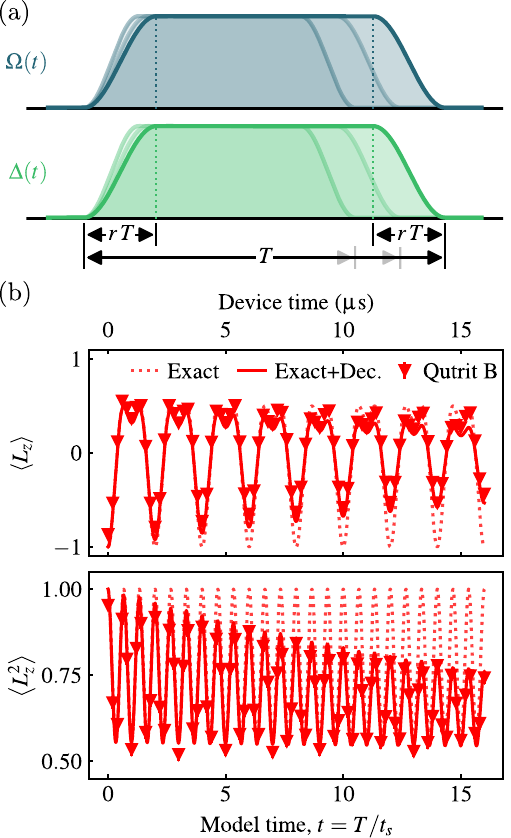}%
\begin{subfigure}{0\linewidth}
\caption{}\label{fig:analog_protocol_single}
\end{subfigure}%
\begin{subfigure}{0\linewidth}
\caption{}\label{fig:analog_data_single}
\end{subfigure}%
\caption{Single qutrit analog experiment.
(a) Protocol for performing a single-qutrit analog simulation of the Abelian Higgs model. Drives on both transitions of the qutrit follow the same envelope $\Omega_{01}(t)=\lambda\Omega_{12}(t)=\Omega(t)=\Omega_0 A_T(t)$ and have equal and opposite detunings $\Delta_1(t) = -\Delta_2(t) =\Delta(t)=\Delta_0 A_T(t)$. Each data point in \subref{fig:analog_data_single} is measured by setting a pulse duration $T$ and reading out qutrit eigenstate populations at the end of the pulse. The ramps are a fixed fraction $r=0.1$ of the total duration, producing a stretching of the pulse for increasing $T$. The dimensionless envelope function $A_T(t)$ is normalized to have an area equal to its duration $T$; see Appendix~\ref{app:rotating_frame} for details. (b) Expectation values of field operators $L_z$ (top) and $L_z^2$ (bottom) for a single qutrit under analog evolution starting in initial transmon state $\ket{n=2}$, with $\kappa/2\pi=\chi/2\pi=1$, $\beta=0$. The scale frequency was chosen to be $t_s^{-1}=\qty{1}{\MHz}$ and converts between physical device time (upper axis) and dimensionless model time (lower axis). Dotted lines show exact Schr\"odinger equation evolution under the model Hamiltonian, solid lines are master equation simulations assuming the independently measured dephasing and relaxation times of Qutrit B in Table~\ref{tab:analog_coherence}.  Uncertainties are estimated from shot noise $\sigma\sim 1/\sqrt{N_\mathrm{shots}}$ where $N_\mathrm{shots}=3000$.
}
\label{fig:analog_single}
\end{figure}

It is possible to map Eq.~\eqref{eq:H_rwa} to Eq.~\eqref{eq:single_qutrit_model} by assuming equal drive phases $\phi_1=\phi_2=-\pi/2$, constant drive envelopes $\Omega_1(t)=\lambda\Omega_2(t)=\Omega_0$, and detunings $\Delta_{1}=-\Delta_{2}=\Delta_0$, which simplifies the driven transmon Hamiltonian to,
\begin{equation}
\begin{aligned}
    H_k^\mathrm{DF} & \approx -\Delta_0 \ketbra{1}{1} \\
    &\phantom{\approx\ } -\frac{1}{2}\Omega_0 \left(\ketbra{1}{0} + \ketbra{2}{1} + \mathrm{h.c.}\right) \\
    & \rightarrow \Delta_0 L_z^2 - \frac{\Omega_0}{\sqrt{2}}L_x,\label{eq:single_qutrit_drive_static}
\end{aligned}
\end{equation}
where in the last step we have removed a global energy offset $-\Delta_0 I$. To convert between the dimensionless parameters in Eq.~\eqref{eq:single_qutrit_model} and the physical parameters of Eq.~\eqref{eq:single_qutrit_drive_static}, we must choose a time scale $t_s$, which sets a scale frequency $t_s^{-1}$. Then the mapping is complete with $\Omega_0 = \chi\,t_s^{-1}$ and $\Delta_0 = \frac{\kappa+\beta}{2} t_s^{-1}$. We have freedom in the single-qutrit experiment to choose the scale frequency as long as the RWA conditions are met. In the experiment shown in Fig.~\ref{fig:analog_single}, we chose $t_s^{-1}=\qty{1}{\MHz}$.

While the above derivation assumed constant drive envelopes in the final step, the experiment must be performed with a shaped envelope, either explicitly or implicitly. To do so in a controlled manner, we give both drives the same envelope: $\Omega_1(t)=\lambda\Omega_2(t)=\Omega_0 A_T(t)$, where $A_T(t)$ is normalized such that its integral after a pulse of duration $T$ is equal to the area of a unit step of the same duration: $\intf{0}{T}{A_T(t)}{t} = T$. The envelope function and its normalization are given explicitly in Appendix~\ref{app:rotating_frame}. The unitary evolution produced at the end of the drive can be modeled with average Hamiltonian theory \cite{brinkmann2016introduction} (see also Appendix~\ref{app_AHT}), where, to first order, the average Hamiltonian is provided by independent time-averages of the individual matrix elements to produce the same result as in Eq.~\eqref{eq:single_qutrit_drive_static}.

Second-order and higher terms in the average Hamiltonian produce coherent errors in the simulation. Because they depend on the commutator of the Hamiltonian with itself at different times, we chirp the pulses during the ramps to maintain instantaneous detunings with the same envelope as the drives, that is, $\Delta_{01}(t)=-\Delta_{12}(t)=\Delta_0 A_T(t)$. Details of the phase ramping necessary to maintain this relationship and the accompanying frame transformation can be found in Appendix~\ref{app:rotating_frame}. The single qutrit Hamiltonian may then be written,
\begin{equation}
\begin{aligned}
    H_{k}^\mathrm{DF}(t) & \approx \Delta(t) L_z^2 - \frac{\Omega(t)}{\sqrt{2}}L_x\\
    & \approx A_T(t)\left(\Delta_0 L_z^2 - \frac{\Omega_0}{\sqrt{2}}L_x\right),\label{eq:singlequtritdrivewithconditions}
\end{aligned}
\end{equation}
and the Magnus expansion then terminates after the first-order term.

As is illustrated in Fig.~\ref{fig:analog_protocol_single}, to simulate evolution to different times, the pulse duration $T$ is varied, and the drive envelope is stretched with a fixed amplitude and a ramp that is a fixed fraction $r=0.1$ of the pulse duration. Then, regardless of $T$, the average Hamiltonian at the end of the pulse is
\begin{equation}
\bar{H}_k^\mathrm{DF} \approx \Delta_0 L_z^2 - \frac{\Omega_0}{\sqrt{2}}L_x,\label{eq:single_avg_hamiltonian}
\end{equation}
where the approximation is due only to the RWA.

The results of a single-transmon analog simulation experiment may be seen in Fig.~\ref{fig:analog_data_single}, where we plot expectation values of the diagonal field operators $L_z$ and $L_z^2$, which are computed directly from measured transmon eigenstate populations, after a statistical readout correction has been applied \cite{bravyi_mitigating_2021,wang_high-e_je_c_2025}. \revision{As is described in Sec.~\ref{sec:model}, these observables are analogous to the local electric field and electric field energy at this site.} For this measurement, the parameters of Eq.~\eqref{eq:single_qutrit_model} were set to $\kappa/2\pi=\chi/2\pi=1$, $\beta=0$. The exact single-qutrit dynamics predicted by the model are plotted along with the data, as well as a Lindblad master equation model described in Appendix~\ref{app:analog_coherence_time_model} that takes into account independently measured decoherence and relaxation times for this transmon, as listed in Table~\ref{tab:analog_coherence} (Qutrit B). We see good agreement with the prediction of the Lindblad model, indicating that the errors in our single-transmon simulation are dominated by decoherence.

\subsection{Two-qutrit implementation}

\subsubsection{Engineering the interaction Hamiltonian}
We next discuss the analog-digital method for engineering the target interaction.
The interaction part of the Hamiltonian in Eq.~\eqref{eqn_latticeModel} takes the form of an Ising-type interaction,
\begin{equation}
\begin{aligned}
    H_\mathrm{AHM,int} = -\beta L_z \otimes L_z.\label{eq:interaction_target}
\end{aligned}
\end{equation}
Because we are performing the simulation on two sites only, we will indicate which site an operator acts on by where it falls relative to the $\otimes$ symbol rather than a superscript index. In addition, when the meaning is clear from context, we will omit the $\otimes$ symbol between operators acting on transmons $A$ and $B$.

Our device consists of a pair of transmons that have a direct capacitive coupling, with properties given in Table~\ref{tab:analog_coherence}. Because our transmon qutrits are far detuned from each other relative to the strength of their capacitive coupling, their native interaction is effectively described by a dispersive \emph{cross-Kerr} Hamiltonian \cite{elliott_designing_2018,malekakhlagh_first-principles_2020}, which is described in more detail in Appendix~\ref{app:interaction_rates}. We parameterize this diagonal interaction into a combination of powers of the $L_z$ operators of the two qutrits,
\begin{equation}
    H_{\mathrm{ck}} = \sum_{i,j=0}^2 z_{ij} L_z^i\sotimes L_z^j,\label{eq:Hck_zzDecomp}
\end{equation}
where $L_z^0 = I$ is the identity matrix. Terms that include $L_z$ to a nonzero power on only one qutrit correspond to local frequency shifts, and may be removed by applying the proper frame rotations to each transmon individually. In the experiment, this corresponds to calibrating the frequencies of each qutrit when the other is in the $\ket{1}$ state, rather than the more common choice of tuning when the other is in the $\ket{0}$ state. This yields a simplified interaction Hamiltonian,
\begin{equation}
\begin{aligned}
    H_{\mathrm{ck}}' &= \sum_{i,j=1}^2 z_{ij} L_z^i\sotimes L_z^j\\
    &=z_{11}L_z L_z + z_{12}L_z L_z^2 \\
    &+ z_{21}L_z^2 L_z + z_{22}L_z^2 L_z^2.\label{eq:zz_coeffs}
\end{aligned}
\end{equation}
We then desire a method that can achieve effective interaction rates $z_{21}=z_{12}=z_{22}=0$.

\begin{figure}[!htb]
\captionsetup[subfigure]{labelformat=nocaption}
\includegraphics[width=\linewidth]{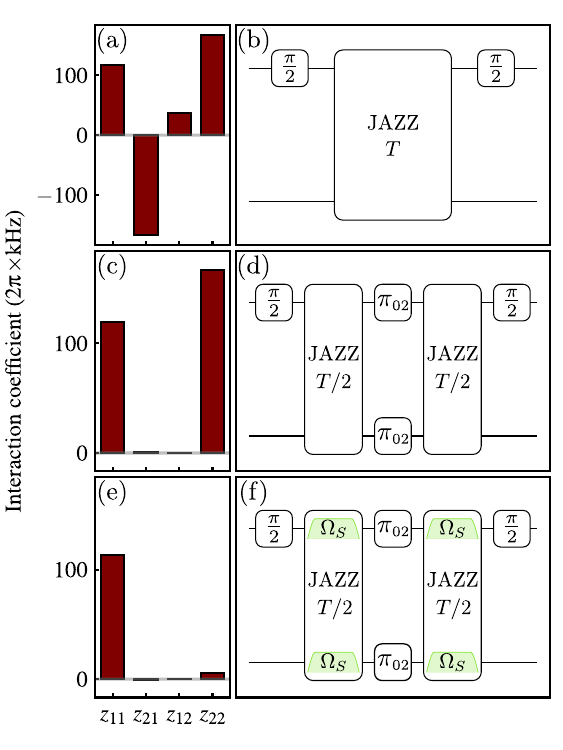}%
\begin{subfigure}{0\linewidth}
\caption{}\label{fig:interaction_rates_bare}
\end{subfigure}%
\begin{subfigure}{0\linewidth}
\caption{}\label{fig:interaction_measurement_bare}
\end{subfigure}%
\begin{subfigure}{0\linewidth}
\caption{}\label{fig:interaction_rates_avg}
\end{subfigure}%
\begin{subfigure}{0\linewidth}
\caption{}\label{fig:interaction_measurement_avg}
\end{subfigure}%
\begin{subfigure}{0\linewidth}
\caption{}\label{fig:interaction_rates_stark_avg}
\end{subfigure}%
\begin{subfigure}{0\linewidth}
\caption{}\label{fig:interaction_measurement_stark_avg}
\end{subfigure}%
\caption{Engineering the $L_z\sotimes L_z$ interaction. (a) Experimentally measured interaction Hamiltonian terms from Eq.~\eqref{eq:Hck_zzDecomp} for the undriven (bare) two-transmon system. (b) Abstract circuit schematic of the interaction rate measurement. The JAZZ block consists of two periods of free evolution of duration $T$, with a $\pi$ pulse on both transmons between them \cite{garbow_bilinear_1982,takita_experimental_2017}. The subspaces of the $\pi$ and $\frac{\pi}{2}$ pulses are determined by which matrix element of the interaction Hamiltonian is being targeted, as is explained in depth in Appendices~\ref{app:interaction_rates} and \ref{app:calibration_sequences}. (c)--(d) Interaction rates and measurement circuit schematic when averaging with a $\ket{0}\leftrightarrow\ket{2}$ pi pulse. We observe that the rates for the odd-parity terms, $z_{12}$ and $z_{21}$ are effectively eliminated by this dynamical decoupling sequence, while the even-parity terms remain unchanged. The circuit now involves four periods of free evolution with duration $T/2$. (e)--(f) Interaction rates and circuit schematic with both averaging and Stark drives applied. Applying Stark drives at the appropriate frequency and amplitude nearly eliminates the driven $\tilde{z}_{22}$ term, leaving the modified $\tilde{z}_{11}$ as the only appreciable interaction rate in the Hamiltonian. See Appendix~\ref{app:calibration_sequences} for details on the calibration of the Stark drives.
}\label{fig:interaction_engineering}
\end{figure}

The values of $z_{ij}$ measured for the transmon pair used in this experiment are shown in Fig.~\ref{fig:interaction_rates_bare}. They are sensitive to the spectra and charge matrix elements of the two transmons, (see Appendix~\ref{app:interaction_rates} for calculations of the rates from second-order perturbation theory), and as a consequence, they cannot be expected in general to realize a native $L_z L_z$ interaction. We measure the rates in the experiment with a double-echo sequence \cite{garbow_bilinear_1982}, referred to as JAZZ in superconducting qubit literature \cite{takita_experimental_2017}, which we have adapted for use with qutrits, (see Appendix~\ref{app:calibration_sequences}).

Undesired interactions may often be canceled using dynamical decoupling techniques \cite{vandersypen_nmr_2005, choi_dynamical_2017, Zhou2024}. However, previously reported pulse sequences on qutrits that can produce the desired interaction involve pulses that do not commute with our single-qutrit drives \cite{Blok:2020may}. Instead, we design a protocol that integrates the single-qutrit drives with a Hamiltonian engineering method based on simultaneous Stark drives on the two transmons and a minimal set of dynamical decoupling pulses, adapted from Ref.~\cite{goss_high-fidelity_2022}. We apply Stark drives to both transmons, $\Omega_{S,A}(t)\cos\omega_S t$ and $\Omega_{S,B}(t)\cos\omega_S t$, where the envelopes $\Omega_{S,k}$ and the frequency $\omega_S$ provide a degree of tunability to the interaction Hamiltonian. This is also the principle behind the controlled-Z gates used in the gate-based simulation in Sec.~\ref{sec:gates}.

In practice, a single pair of Stark drives only affords enough tunability to set one term in the interaction Hamiltonian to zero. With this degree of freedom in mind, we search for sequences consisting only of pulses that leave the single-qutrit Hamiltonians invariant and that may lead to a cancellation of at least two of the undesired terms.  Conveniently, the operation $U_{02}\equiv\pi_{02}^A\pi_{02}^B$, where $\pi_{02}^k$ is a local unitary that transposes states $\ket{0}$ and $\ket{2}$ on transmon $k$, commutes with the single-transmon average Hamiltonian in Eq.~\eqref{eq:single_avg_hamiltonian} and has the effect of flipping the signs on the odd-parity interaction terms while leaving the even-parity terms unchanged:
\begin{equation}
\begin{aligned}
    U_{02}H_{\mathrm{ck}}'U_{02}^\dagger &=z_{11}L_z L_z - z_{12}L_z L_z^2 \\
    &- z_{21}L_z^2 L_z + z_{22}L_z^2 L_z^2.
\end{aligned}
\end{equation}

By alternating periods of evolution under $H_{\mathrm{ck}}$ with dynamical decoupling pulses $U_{02}$, the system can be made to evolve under the dynamically-decoupled interaction Hamiltonian on average,
\begin{equation}
\begin{aligned}
 H_\mathrm{ck}^{\mathrm{dd}} &= \frac{1}{2}\left(H_{\mathrm{ck}}'+U_{02}^\dagger H_{\mathrm{ck}}'U_{02}\right)\\
 &= z_{11}L_z L_z+z_{22}L_z^2 L_z^2.
\end{aligned}\label{eq:average_interaction}
\end{equation}
As can be seen from average Hamiltonian theory (Appendix~\ref{app_AHT}), as long as the pulses that perform $U_{02}$ are assumed to be instantaneous, this averaging is exact in the absence of other terms in the Hamiltonian, since $H_{\mathrm{ck}}'$ and $U_{02}H_{\mathrm{ck}}'U_{02}^\dagger$ commute. The averaged interaction rates and the protocol for measuring them are shown in Figs.~\ref{fig:interaction_rates_avg} and \ref{fig:interaction_measurement_avg}, respectively.

After this sequence, the only remaining undesired interaction term is $z_{22}L_z^2 L_z^2$, and we find that for a proper choice of Stark drive frequency and amplitude, the coefficient $z_{22}\rightarrow \tilde{z}_{22}$ can be driven close to zero. This modifies all the coefficients in the driven interaction Hamiltonian, $z_{ij}\rightarrow\tilde{z}_{ij}$, and we refer to the new effective cross-Kerr interaction Hamiltonian as $\tilde{H}_\mathrm{ck}'$ and the effective driven and dynamically-decoupled interaction as $\tilde{H}_\mathrm{ck}^\mathrm{dd}$. The measured coefficients of $\tilde{H}_\mathrm{ck}^\mathrm{dd}$ are shown in Fig.~\ref{fig:interaction_rates_stark_avg}. We observe that only the $L_z L_z$ term has appreciable weight after applying the protocol. The calibration procedures for finding the Stark drive frequency and amplitudes required to achieve $\tilde{z}_{22}\approx 0$ are described in Appendices~\ref{app:interaction_rates} and \ref{app:calibration_sequences}, respectively.

While the single-qutrit experiment has great flexibility in the choice of scale frequency, in the two-qutrit experiment $\tilde{z}_{11}$ is fixed after applying the Stark drives, and thus it sets the scale for the other terms in the Hamiltonian. That is, we have $\tilde{z}_{11}=-\beta\,t_{s}^{-1}$ with $\beta/2\pi\equiv 1$. The minus sign appears
due to a mismatch in signs between $\tilde{z}_{11}>0$ and the $-\beta L_z L_z$ term in the model Hamiltonian.
We account for this by reversing the mappings of transmon states $\ket{n=0}$ and $\ket{n=2}$ in Eq.~\eqref{eq:qutrit_mapping} for only one transmon, effectively conjugating its entire evolution with $\pi_{02}$, since $L_z\sotimes \pi_{02}^\dagger L_z \pi_{02} = -L_z\sotimes L_z$. This is done entirely in post-processing.

\subsubsection{Combining on-site and interaction terms}

\begin{figure}[!htb]
\captionsetup[subfigure]{labelformat=nocaption}
\includegraphics[width=0.9\linewidth]{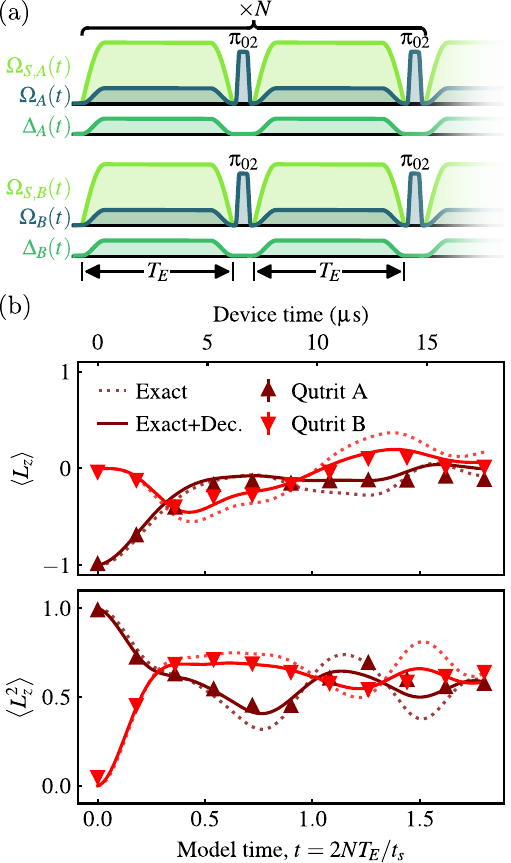}%
\begin{subfigure}{0\linewidth}
\caption{}\label{fig:analog_protocol_double}
\end{subfigure}%
\begin{subfigure}{0\linewidth}
\caption{}\label{fig:analog_data_double}
\end{subfigure}
\caption{Two qutrit analog-digital experiment.
(a) Two-qutrit protocol. Evolution of the simulation is achieved by tiling evolution periods of a fixed duration $T_E=\qty{800}{\ns}$ interleaved with \qty{60}{\ns} $\pi_{02}$ pulses on both transmons. Data points in \subref{fig:analog_data_double} are measured after $2N$ evolution periods. The $\pi_{02}$ pulses are performed by driving both $\ket{0}\leftrightarrow\ket{1}$ and $\ket{1}\leftrightarrow\ket{2}$ transitions on resonance simultaneously \cite{forney_multifrequency_2010,champion_multi-frequency_2024}. During each evolution period, in addition to the detuned drives described in Sec.~\ref{sec:single_qutrit_analog}, a Stark tone $\Omega_S(t)$ is applied to each transmon, which modifies the entangling Hamiltonian $\tilde{H}_\mathrm{ck}$ such that the average interaction after two steps is approximately $\tilde{H}_\mathrm{ck}^\mathrm{dd}\approx\tilde{z}_{11}L_z \sotimes L_z$. (b) Expectation values of $L_z$ (top) and $L_z^2$ (bottom) for two qutrits during the analog-digital protocol. Dotted lines show exact theory evolution, solid lines are master equation simulations assuming independently measured dephasing and relaxation times. The scale frequency was fixed so that $t_s^{-1}=\tilde{z}_{11}/2\pi\approx\qty{110}{\kHz}$. The initial transmon state was $\ket{\psi_\mathrm{init}}_{\!AB}=\ket{21}$.}
\label{fig:analog_double}
\end{figure}

We combine the single-qutrit drive Hamiltonians $H_k^\mathrm{DF}(t)$ defined in Sec.~\ref{sec:single_qutrit_analog} with the Stark-driven interaction Hamiltonian by turning on all the drives simultaneously, as sketched for each pulse in Fig.~\ref{fig:analog_protocol_double}, yielding
\begin{equation}
H_\mathrm{sys}(t) = H_A^\mathrm{DF}(t) + H_B^\mathrm{DF}(t) + \tilde{H}_\mathrm{ck}'(t),\label{eq:analog_system_hamiltonian}
\end{equation}
for $t\in[0,T_E]$, where $T_E$ is the fixed evolution period for a single pulse.
To reduce errors caused by a failure of the Hamiltonian to commute with itself at different times throughout the pulse, we must ensure that the time dependence of $\tilde{H}_\mathrm{ck}'(t)$, determined by the product $\Omega_{S,A}(t)\Omega_{S,B}(t)$, varies with the same envelope as the single-qutrit drives and detunings. Hence, we set the Stark drive envelopes to be $\Omega_{S,A}(t) = \Omega_{S,B}(t)=\Omega_S^0\sqrt{A_{T_E}(t)}$. This does not entirely eliminate the error, as there remains the undriven component of the interaction Hamiltonian, which does not commute with the single-qutrit $L_x$ terms. As we show in Appendix~\ref{app:analog_error_terms}, however, this error only enters to third order in the average Hamiltonian and scales roughly as $(rT_E)^2$, quadratically with the ramp duration.

We cannot reduce the ramp duration indefinitely, since we find in practice that the ramp must also be sufficiently long to maintain an adiabatic change of the interaction Hamiltonian. In this experiment, we chose a ramp duration of \qty{80}{\ns}. When the errors are made sufficiently small, the effective average Hamiltonian over the course of one pulse is well approximated by the first order term in the Magnus expansion,
\begin{equation}
\begin{aligned}
    \bar{H}_\mathrm{sys} &\approx \frac{1}{T_E}\intf{0}{T_E}{H_\mathrm{sys}(t)}{t}\\
    &= \bar{H}_A^\mathrm{DF} + \bar{H}_B^\mathrm{DF} + \bar{\tilde{H}}_\mathrm{ck}'.\label{eq:full_analog_first_order}
\end{aligned}
\end{equation}

Then alternating between periods of driven evolution and the $\pi_{02}$ pulses---which in this experiment are performed by driving both $\ket{0}\leftrightarrow\ket{1}$ and $\ket{1}\leftrightarrow\ket{2}$ transitions simultaneously with zero detuning at a much larger amplitude than the on-site simulation drives \cite{forney_multifrequency_2010,champion_multi-frequency_2024}---we arrive at the desired effective Floquet evolution:
\begin{equation}
\begin{aligned}
    U_\mathrm{eff}(N)&=\left[\exp\!\left(-i U_{02}^\dagger\bar{H}_\mathrm{sys}U_{02}T_E\right)\exp\!\left(-i \bar{H}_\mathrm{sys}T_E\right)\right]^N\\
    &\approx \exp\!\left(-2i \left(\bar{H}_A^\mathrm{DF} + \bar{H}_B^\mathrm{DF} + \tilde{H}_\mathrm{ck}^\mathrm{dd}\right)NT_E\right),
\end{aligned}
\end{equation}
where $\bar{H}_A^\mathrm{DF}$ and $\bar{H}_B^\mathrm{DF}$ are given by Eq.~\eqref{eq:single_avg_hamiltonian} and $\tilde{H}_\mathrm{ck}^\mathrm{dd}\approx \tilde{z}_{11}L_z L_z$. This therefore approximates $H_\mathrm{AHM}$.

The error in the approximation can be thought of in terms of both the higher-order terms in the Magnus expansion and Trotterization error. We analyze the error terms using AHT in Appendix~\ref{app:analog_error_terms} and find that they scale linearly in the evolution period $T_E$, suggesting that shorter evolution periods between the $\pi_{02}$ gates will produce a better approximation. There is a tradeoff, however, with the requirement that the gates be effectively instantaneous compared with $T_E$. In practice, we find an acceptable balance with an evolution pulse duration $T_E=\qty{800}{\ns}$ and a $\pi_{02}$ gate time of $T_g=\qty{60}{\ns}$.

\subsubsection{Device results}
In Fig.~\ref{fig:analog_data_double} we plot dynamics from the experimental implementation of the analog-digital protocol, along with numerically integrated evolution \cite{lambert2024qutip5quantumtoolbox} of the Hamiltonian in Eq.~\eqref{eqn_latticeModel}, as well as a dissipative model described by a Lindbladian that takes into account relaxation and dephasing times (Table~\ref{tab:analog_coherence}) of the transmons used in the experiment. We find that the model with decoherence captures the dynamics of the driven system well.
To account for the effect of periodic population flips in the decoherence model with a time-independent master equation, we assume an effective averaged Lindblad equation, where we take the jump operators to be transformed by the periodic application of $\pi_{02}$ pulses in the same manner as the Hamiltonian.
This effective Lindblad model is described in detail in Appendix~\ref{app:analog_coherence_time_model}. We also plot the transmon eigenstate populations for multiple initial product states along with the predictions of the model in Fig.~\ref{fig:analog_double_extended} in the appendix.

\section{Gate-based Simulation} \label{sec:gates}

\begin{figure*}[!t]
    \begin{centering}
    \subfloat{\includegraphics[width=0.45\linewidth]{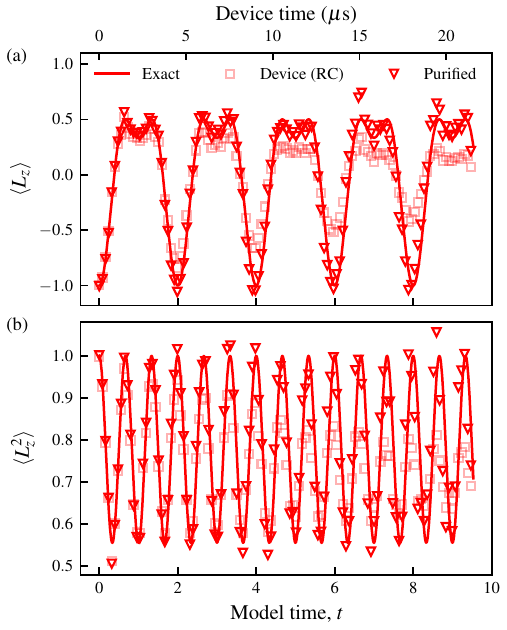 }\label{fig_1Q_lz} }~
    \subfloat{\includegraphics[width=0.45\linewidth]{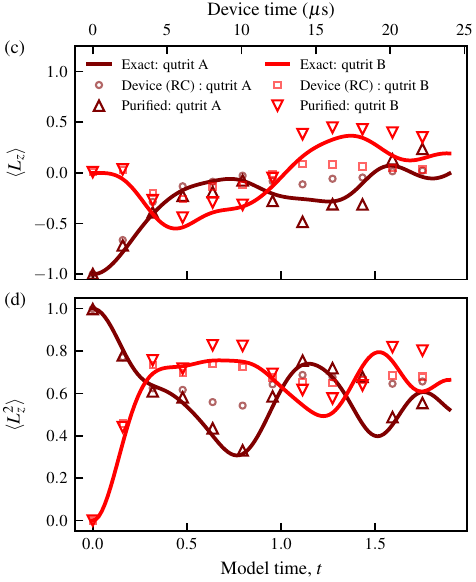} \label{fig_2Q_lz} }~
    \end{centering}
    \caption{Exact diagonalization (Exact), device data with randomized compilation (Device (RC) ) and error-mitigated device results (Purified) of field operators $\langle L_{z}\rangle$ and $\langle L_{z}^{2} \rangle$ at (a,b) $\kappa/2\pi=\chi/2\pi=1.0$, $2\pi\delta t = 0.55$ for the single qutrit experiment and (c,d) $\kappa/2\pi=\beta/2\pi=\chi/2\pi=1.0$, $2\pi\delta t = 1.0$ for the two-qutrit experiment. For a discussion on the impact of different types of error mitigation schemes see Appendix~\ref{app_digital_mitigation}. The dynamics start from the initial state $\ket{\psi_\mathrm{init}}_{\!AB}=\ket{21}$.}
    \label{fig_2Q_lz_lzsqr}
\end{figure*}

In this section, we describe the digital gate-based method for simulating the AHM and present the dynamics obtained in experiment. We perform the digital simulation on a Berkeley-based qutrit processor, the properties of which are listed in Table~\ref{table_Berkeley} in Appendix~\ref{app_gate_approx}. The discussion in this section focuses on the mapping of the evolution produced by the AHM to the native gate set of our processor. We also highlight the importance of error mitigation schemes in qudit simulations for obtaining reliable results with present-day hardware and discuss the associated computational cost.

\subsubsection{Mapping of the unitary operations to basis gates}
For an $N_s$-qutrit system, the time evolution operator under the model Hamiltonian in Eq.~\eqref{eqn_latticeModel} can be approximated using a first-order Trotter approximation,
\begin{equation}
\begin{aligned}
 &U(t)= \exp(-iH_\mathrm{AHM}t) \\
 &\simeq \Big( \prod_{k=1}^{N_s} M^{k}_z (\kappa,\delta t)
\prod_{k=1}^{N_s-1} M_{zz}^{k,k+1} (\beta, \delta t)
\prod_{k=1}^{N_s} M_x^{k} (\chi,\delta t)
\Big)^n,
\end{aligned}\label{eq:trotter_unitary}
\end{equation}
where the evolution time $t$ is discretized into uniform steps $\delta t$ such that  $t= n \,\delta t$. Here,  $M_z^{k}=\exp(-i\theta (L_z^{(k)})^{2})$ and $M_x^{k}=\exp(-i\phi L_x^{(k)})$ denote on-site operations that act on an individual qutrit $k$, while $M_{zz}^{k,l}=\exp(-i\gamma L_z^{(k)} L_z^{(l)})$ represents an entangling operation between a pair of qutrits $k$ and $l$. The angles $\phi$, $\theta$, and $\gamma$ are determined by the Hamiltonian parameters $\chi$, $\kappa$, and $\beta$, as well as the Trotter step duration $\delta t$. The matrix representations of these operations are presented in Appendix~\ref{app_gate_approx}. The error in the approximation is bounded by $\mathcal{O}(t \,\delta t)$. An improved error scaling can be achieved with a higher order Trotter approximation with an increased cost in the number of entangling gates.

To perform the operations $M_z^k$, $M_x^k$, and $M_{zz}^{k,l}$ on our superconducting qutrit processor requires further decomposition of the unitaries into our native gate set. In contrast to the multi-tone drives used for the analog-digital simulation, we here use single-tone drives that are resonant with a single transition on each transmon qutrit. Such drives can achieve qubit-like rotations in the subspace spanned by either the states $\{\ket{0},\ket{1}\}$ or $\{\ket{1},\ket{2}\}$ \cite{Morvan:2020rrb}. The fixed subspace rotations $\{R_{x}^{01}(\frac{\pi}{2}),R_{x}^{12} (\frac{\pi}{2}),R_{y}^{01}(\frac{\pi}{2}),R_{y}^{12} (\frac{\pi}{2})\}$ and arbitrary phase gates, $R_z^{01}(\theta)$ and $R_z^{12}(\theta)$, which are performed by adjusting the phases of later microwave drives \cite{mckay_efficient_2017}, then constitute the native gate set for each qutrit, where we define
\begin{equation}
    R_\alpha^{ij}(\theta)=\exp\!\left(-i\frac{\theta}{2} \sigma_\alpha^{ij}\right),\quad \alpha\in\{x,y,z\}.
\end{equation}
The $\sigma_\alpha^{ij}$ are the generators of qubit-like rotations around axis $\alpha$ within the $\{\ket{i},\ket{j}\}$ subspace of the qutrit Hilbert space and can be expressed in terms of the Gell-Mann matrices (see Appendix~\ref{app_gate_approx}).
Then, in combination with the controlled-\textsc{sum} entangling gate $U_\mathrm{CSUM}\ket{i,\,j}=\ket{i,\,j+i\pmod{d}}$ we have a gate set that is sufficient for universal qutrit computation \cite{gottesman_fault-tolerant_1999, di_synthesis_2013}.

To represent an arbitrary single qutrit operation $\exp(i \, h)$ using the elementary gates, we first expand $h$ in the Gell-Mann basis $\{\lambda_i\}$, $h=\sum_{k=1}^8 c_k \lambda_k$, where the coefficients are computed with $c_k=\frac{1}{2}\mathrm{Tr}(h\lambda_k)$. Then we perform a first-order Trotter approximation by exponentiating each term in the sum separately. For the $M_z^k$ term this yields an exact decomposition (up to a global phase),
\begin{equation}
M_z (\theta) =  R_z^{01}\!\left(\frac{-2\theta}{3}\right)R_z^{12}\!\left(\frac{2\theta}{3}\right), \label{eqn_Mz}
\end{equation}
where $\theta= (\frac{\kappa}{2} +\beta) \delta t$ for all qutrits except for the boundary qutrits, where instead $\theta=\frac{\kappa+\beta}{2}\delta t$. For the case of $N_s=2$ studied here, both qutrits are treated as boundary qutrits.

For the $M_x^k$ term, we arrive at an approximate result,
\begin{equation}
    M_x (\phi)\simeq
    R_x^{01}\!\left(\sqrt{2}\phi\right)R_x^{12}\!\left(\sqrt{2}\phi\right), \label{eqn_Mx}
\end{equation}
where $\phi=\frac{\chi}{\sqrt{2}} \delta t $.
We note that the Trotter approximation in generating the $M_x$ operation could be removed by adding a calibrated $\pi/2$ spin-1 rotation to our digital gate set by applying multitone drives \cite{Zhou2024,champion_multi-frequency_2024}, but this is left for future work.

Following a similar analysis, a decomposition for the entangling operation $M_{zz}$ can be found by expressing the two-qutrit Hermitian operator in terms of tensor products of Gell-Mann matrices,  $h = \sum_{k,l} c_{kl} \lambda_k \sotimes \lambda_l$. This results in four non-zero coefficients. A much simpler decomposition, however, is found by inspection:
We decompose the entangling operation $M_{zz}$ into three controlled-\textsc{sum} gates interleaved with local phase gates,
\begin{align}
M_{zz} (\gamma) =U_\mathrm{CSUM}\, D(\gamma)\, U_\mathrm{CSUM}\, G(\gamma)\, U_\mathrm{CSUM}\label{eqn_ent_gate}
\end{align}
where $\gamma=\delta t\, \beta $. $D$ and $G$ are diagonal matrices constructed from the elementary phase gates,
\begin{align}
    D (\gamma)&=  R_z^{01}\Big(\frac{4\gamma}{3} \Big) R_z^{12}\Big(\frac{2\gamma}{3}\Big),\\
    G (\gamma)&=  R_z^{01}\Big(\frac{2\gamma}{3} \Big) R_z^{12}\Big(\frac{4\gamma}{3} \Big).
\end{align}
In addition to our circuit decomposition analysis above, we tested whether a more efficient compilation was possible with numerical synthesis methods provided by the TrueQ software \cite{beale2020true}. We found comparable costs in entangling operations from numerical synthesis to that of our analytical decomposition. For completeness and a fair comparison with future experiments, a complete set of native gates of the AQT testbed and the key device parameters such as qutrit frequencies, anharmonicity, relaxation time and dephasing times are listed in Appendix~\ref{app_gate_approx}.

The digital simulation provides a more direct way to compare against a hypothetical digital qubit-based simulation of the same model. In addition to the need for two qubits to encode the state of each three-level site, we find that around 14 entangling operations are required in a qubit-based implementation with all-to-all connectivity demonstrated in Appendix~\ref{app_qubit_resource_estimate}, which is nearly a fivefold increase compared to the three native entangling operations required in the digital qutrit protocol. The gate depth would need to increase further for a processor with restrictive connectivity.  We find with compilations based on IBM's heavy-hex topology that the added requirement of SWAP operations would increase the gate depth by roughly another factor of ten.

Thus, if the gate times for the native entangling operations, relaxation times, and dephasing times of transmon qutrit processors become comparable to state-of-the-art qubit processors, qutrit-based implementations can be favorable for the implementation for certain gauge theory models that inherit $\mathrm{SU}(3)$ group structure. This is particularly relevant for simulating various lattice gauge theories since $\mathrm{SU}(3)$ appears  in different limits of QCD \footnote{SU(3) appears as the color gauge group of QCD and in the low energy limit applying massless quark approximation we get an effective low energy theory that has global left-right symmetry $\mathrm{SU}(3)_L\times \mathrm{SU}(3)_R$ which is spontaneously broken to $\mathrm{SU}(3)$ vector symmetry due to chiral condensate.}. Thus the phase structure of QCD and variants of QCD with various degrees of approximation can likely be studied more efficiently with digital qutrit-based quantum processors \cite{Ciavarella:2021nmj,
gustafson_prospects_2021,jiang_non-abelian_2025}.

\subsubsection{Device results}
We now present hardware results of our gate-based simulation for the single-site and two-site lattice Abelian Higgs model. In the single-site experiment, the fast high-fidelity single-qutrit gates allow us to observe dynamics for longer times compared to the two-site experiment.
Fig.~\ref{fig_2Q_lz_lzsqr} shows error-mitigated hardware results for \revision{the value of the field,} $\langle L_z^{(k)} \rangle$, and \revision{local field energy,} $\langle (L_z^{(k)})^{2} \rangle$ expectation values, which are calculated from the measured transmon eigenstate populations, compared with the exact diagonalization and the hardware results with randomized compilation. %
We find improved fidelity of the circuit operation as we apply qudit randomized compiling \cite{Magesan:2010tmc,Magesan:2012xfe, hashimrc, Goss:2023frd} to mitigate coherent errors and rescale our expectation values with noise measured from cycle benchmarking \cite{Erhard:2019cxk,Carignan-Dugas:2023fgt}.

We use 30 distinct twirled versions of the Trotterized circuit where  we use $N_{\mathrm{shots}}=1024$ shots per twirled circuit. Twirling refers to applying local random single-qutrit Pauli operations for each qutrit before and after each cycle of entangling operation. The random operations are selected from the set of generalized Paulis which form an orthogonal basis for a single qutrit. We also employ cycle benchmarking to learn the error rates of different Pauli and SPAM channels associated with the native gates in our circuit, allowing us to correct the observables by post-processing the estimate with a numerical correction factor. Additionally, readout-error mitigation is performed by confusion matrix inversion.
The purpose of twirling is to transform coherent noise to stochastic noise channels, which improves the scaling for the accumulation of errors-from quadratic to linear order in circuit depth. This also makes the subsequent mitigation step more effective, as state purification is used to mitigate the incoherent part of the device noise learned from our cycle benchmarking experiments. For a quantitative analysis on the effectiveness of the mitigation schemes, readers are advised to consult the discussion in Appendix~\ref{app_digital_mitigation}.

\section{Discussion and Outlook}\label{sec:interpretation}
\begin{figure}[b]
    \centering
    \subfloat{%
        \includegraphics[width=0.46\textwidth]{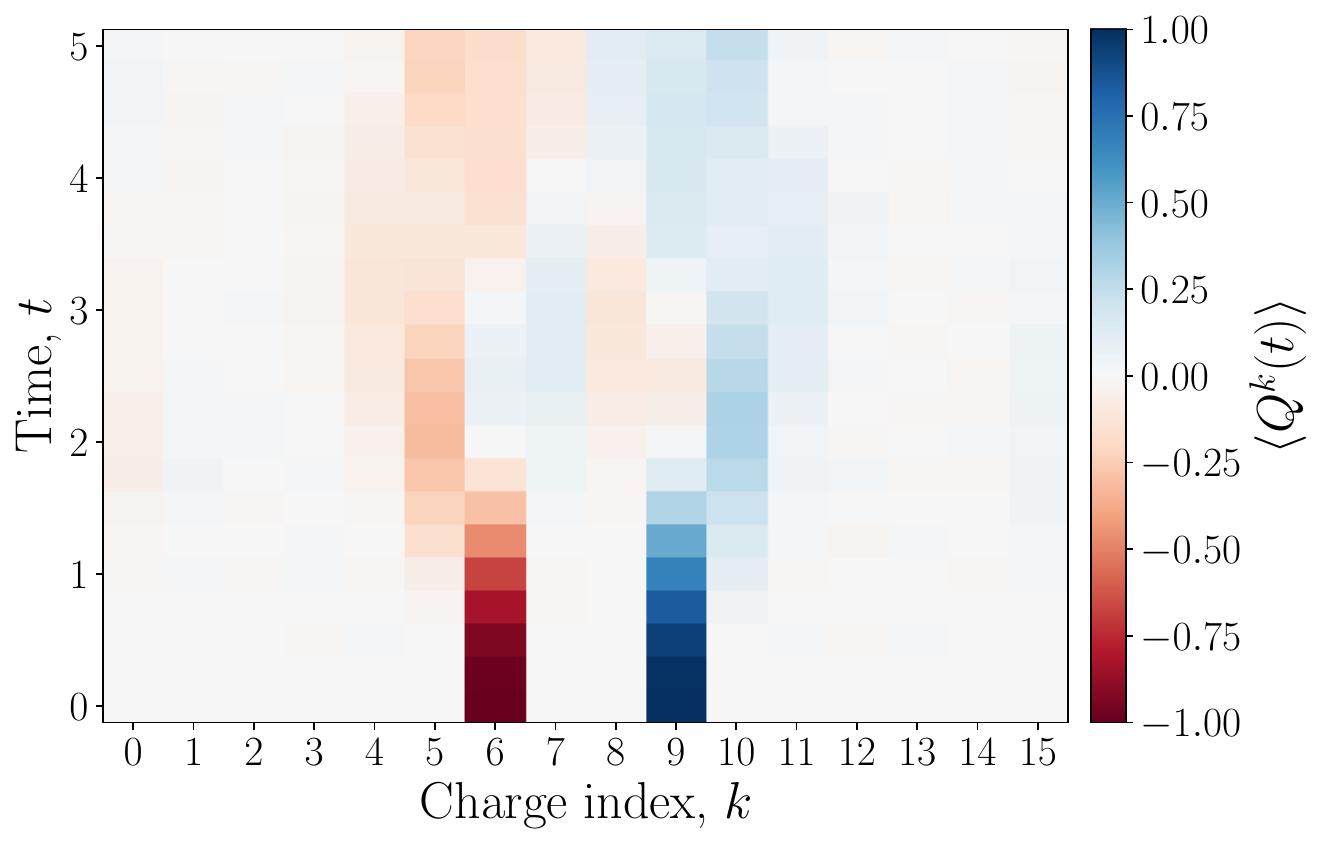}
    }
    \caption{Demonstration of scalability of the digital simulation protocol with a noiseless simulator \cite{Cirq_Developers_2025}. Here we show results of a \emph{Cirq} simulation of the Trotterized unitary in Eq.~\eqref{eq:trotter_unitary} for a the 15-site lattice model with $\kappa/2\pi=\beta/2\pi=\chi/2\pi=1.0$. The initial state with two domain walls $|\psi_\mathrm{init}\rangle=|111111000111111 \rangle$ was prepared and the time-dependence of local charge is computed as the difference in $\langle L_{(z)}\rangle$ between adjacent qutrits: $\langle Q^k(t)\rangle = \langle L_{(z)}^{(k-1)}\rangle - \langle L_z^{(k)}\rangle$. For open boundary conditions, additional edge sites are assumed to be fixed at $\langle L_z^{(-1)}\rangle=\langle L_z^{(15)}\rangle=0$ when computing $\langle Q^k\rangle$.}
\label{fig_breaking}
\end{figure}

\begin{figure*}[!htb]
\captionsetup[subfigure]{labelformat=nocaption}
\includegraphics[width=0.9\linewidth]{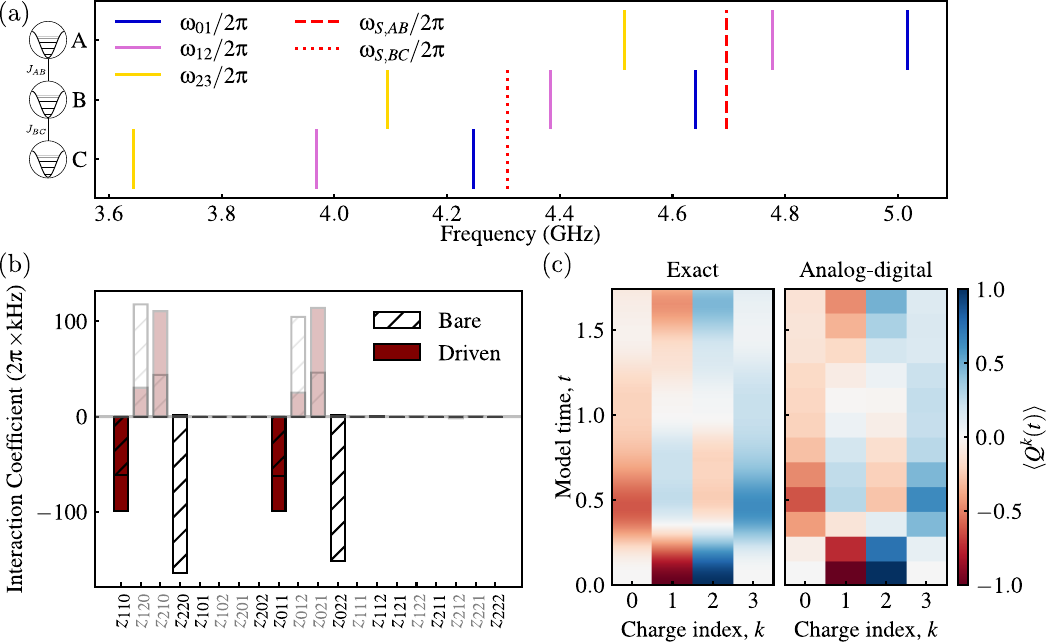}%
\begin{subfigure}{0\linewidth}
\caption{}\label{fig:triple_chip_properties}
\end{subfigure}%
\begin{subfigure}{0\linewidth}
\caption{}\label{fig:interaction_rates_triple}
\end{subfigure}%
\begin{subfigure}{0\linewidth}
\caption{}\label{fig:analog_sim_triple}
\end{subfigure}
\caption{Demonstrating the feasibility of scaling up the analog-digital protocol with numerical simulations.
(a) Transition frequencies of the three simulated transmons, along with the Stark drive frequencies used to engineer the desired interactions. Note that the transmon-transmon detunings are larger here than in the experiment. Capacitive couplings $J_{kl}$ were set to be $J_{AB}=\qty{2.42}{\MHz}$, $J_{BC}=\qty{2.40}{\MHz}$. (b) Interaction rates of the chain of transmons. In the Bare case (empty hatch), coefficients $z_{ijk}$ are computed from numerical diagonalization of the three-transmon Hamiltonian $H = \sum_{ijk\in\{0,1,2\}} z_{ijk} L_z^i\sotimes L_z^j\sotimes L_z^k$, where superscripts indicate powers of the $L_z$ operator. In the driven case (maroon), $\tilde{z}_{ijk}$ are calculated by fitting to the phases accumulated after numerically integrating the Schr\"odinger equation while Stark drives are applied to both pairs of transmons simultaneously. Odd-parity terms under conjugation with $U_{02}$ are grayed out to indicate that they will be canceled by the dynamical decoupling protocol, leaving only $\tilde{z}_{110}/2\pi=\qty{-99.3}{\kHz}$ and $\tilde{z}_{011}/2\pi=\qty{-99.4}{\kHz}$. (c) Numerically simulated evolution of the analog-digital simulation of the AHM on three noiseless transmons with model parameters $\kappa/2\pi=\chi/2\pi=\beta/2\pi=1$ and a scale frequency $t_s^{-1}=\frac{1}{2\pi}(\tilde{z}_{011}+\tilde{z}_{110})/2$, compared to exact evolution under the AHM. Local charge $\langle Q^k\rangle$ is calculated as the difference in $\langle L_z\rangle$ between adjacent qutrits as in Fig.~\ref{fig_breaking}.
The initial transmon state was $\ket{\psi_\mathrm{init}}=\ket{101}$.
}
\label{fig:analog_triple}
\end{figure*}

The analog-digital and fully digital approaches to simulating the Abelian Higgs model presented here each have their own potential advantages and remaining challenges. The main strength of performing the simulation on a digital qudit processor is that it is universal, allowing for great freedom in varying the model parameters, and in exploring related models, as long as a circuit decomposition may be found. At the same time, the native gate set of the processor may not always permit efficient decompositions of the target operations, as we see here in the need for the Trotterized approximation of the $M_x$ operation and the requirement of three controlled-\textsc{sum} operations per Trotter step.
In contrast, the analog and the analog-digital approaches, while more restrictive in the variety of models they can simulate, demonstrate that it is possible to more directly realize the target Hamiltonian considered here, which was sufficiently close to the device Hamiltonian.
In the remainder of this section, we point out some noteworthy aspects of both protocols, and discuss potential directions for future work.
\begin{table*}[t]
\caption{\label{tab:comparison}%
Summary of the analog-digital and digital protocols
for simulating the spin-1 truncated Abelian Higgs model on transmon qutrits.
Device-independent properties of the protocols are provided in the upper block and device-specific properties are listed in the lower block. Lower block numbers are dependent on the coherence times and gate durations of two distinct processors given in
(Table~\ref{tab:analog_coherence} and Table~\ref{table_Berkeley}) and are provided as a discussion summary rather than a comparative benchmark. More details may be found throughout the main text and the Appendices.}
\begin{ruledtabular}
\begin{tabular}{l >{\raggedright\arraybackslash}p{0.36\textwidth} >{\raggedright\arraybackslash}p{0.36\textwidth}}
 & \textbf{Analog-digital (hybrid)} & \textbf{Digital} \\
\colrule
Time evolution unit
  & Floquet cycle with two analog driven periods and two $\pi_{02}$ pulses
  & First order Trotter implementation of $M_z^{(k)}$, $M_x^{(k)}$, $M_{zz}^{(k,l)}$ operations with native gates \\[2pt]

Entangling operations per step
  & Single Floquet-based engineering of the interaction term
  & $3$ CZ gates \\[2pt]

Control pulses
  & Simultaneous multi-tone and Stark drives, alternating with multi-tone $\pi_{02}$ gates.
  & Single-tone sequential $R^{01}_{x,y}\left(\pm \frac{\pi}{2}\right), R^{12}_{x,y}\left(\pm \frac{\pi}{2}\right)$ pulses, alternating with two-qutrit CZ gates. \\[2pt]

Error mitigation
  & None %
  & Randomized compilation and purification \\[2pt]

Model flexibility
  & Flexible for certain classes of Hamiltonians
  & Universal (model-independent protocol) \\[2pt]

Device restrictions
    & Requires ability to engineer uniform $L_z\!\otimes\! L_z$ rates.
    & Standard qutrit algorithm device design requirements \cite{morvan_optimizing_2022}. \\[2pt]

\colrule
\multicolumn{3}{c}{Device-specific aspects; not a benchmark comparison\footnotemark[1]} \\
\colrule
Platform
  & Rochester qutrit processor
  & Berkeley (AQT) qutrit processor \\[2pt]

Evolution / step timing
  & $T_E = \qty{800}{\ns}$, $\pi_{02}$ gate $= \qty{60}{ns}$
  & CZ $=\qty{612}{\ns}$, SU(3) $1Q =\qty{180}{\ns}$ \\[2pt]

Scale frequency $t_s^{-1}$
  & $\tilde z_{11}/2\pi \approx 110~\mathrm{kHz}$ (two-site)
  & Set by $\delta t$ (variable) \\[2pt]

Control pulse errors
  & $\pi_{02}$ pulse fidelity $\sim\!99.5\%$, imperfect $z_{22}$ cancellation $<\qty{10}{\kHz}$
  & Combined CZ and single-qutrit gate fidelity $\sim\!93\%$ \\

Protocol error\footnotemark[2]
  & Second-order Magnus expansion infidelity: $\sim\!3\%$ per Floquet step
  & Trotter infidelity per Trotter step: $\sim\!5\%$ ($2\pi\delta t = 1.0$), $\sim\!1\%$ ($2\pi\delta t = 0.55$) \\[2pt]

\end{tabular}
\end{ruledtabular}
\footnotetext[1]{Circuit depth and achieved simulation time mix protocol
with the coherence times and gate durations of two different chips and are
therefore not directly comparable across the two columns.}
\footnotetext[2]{Protocol error includes only the contribution to the infidelity due to approximations made in deriving the simulation protocol, assuming ideal control pulses.}

\end{table*}

One way these protocols differ is in the immediate availability of error mitigation methods.
We have demonstrated that the application of Weyl twirling and purification in the digital experiment significantly improves the agreement between the data and the exact diagonalization results. In the analog-digital experiment, however, the same techniques, which rely on the properties of Clifford gates, are not currently available. While there are emerging methods for benchmarking \cite{wilkens_benchmarking_2024, canelles_benchmarking_2023, valahu_benchmarking_2024} and mitigating errors in analog simulations \cite{steckmann_error_2025, guo_mitigating_2025, zemlevskiy_optimization_2024}, they are generally tailored to a particular model or method, and are not yet applicable to the simulation performed here.

When considering scaling up the simulation to longer chains of transmons, techniques for scaling up qubit-based computation can be readily applied to both digital and analog qutrit simulators. Fig.~\ref{fig_breaking} shows numerical simulations of an extension of the digital simulation to a large-scale chain, demonstrating that future experiments with this protocol on near-term qutrit processors may be used to study the dynamics of charges on the lattice and to investigate phenomena such as string breaking and confinement. Scaling up to higher spatial dimensions will introduce additional engineering challenges and gate costs due to the presence of magnetic field terms, which involve four-body interactions.

Scaling the analog-digital protocol, on the other hand, may require a tailor-made device, due to the requirement that all coupled pairs of transmons have the same effective $L_z\sotimes L_z$ interaction rate. We demonstrate in Fig. \ref{fig:analog_triple} with numerical simulations that implementing the (1+1)D AHM \revision{can be done on an idealized chain} of three transmons with one Stark drive per coupled pair, requiring the anharmonicities of the three transmons to be approximately the same, and the detunings between the pairs to also be approximately equal. While this demonstrates that it \revision{may be possible} to scale up the protocol with an appropriately designed device, the frequency spacing requirement is likely too strict for a long chain of fixed-frequency transmons when allowing for fabrication imperfections. To remedy this, one possible avenue is the use of additional Stark drives to modify the interactions further, which we leave for future work.

\revision{Although the analog-digital hybrid approach does impose significant constraints on engineering the device, this simultaneous combination of Stark-modified interaction terms, detuned driving, and dynamical decoupling sequences can be thought of as a new tool in the qudit-based analog simulation toolbox, and may enable the construction of hybrid analog-digital simulations of a more diverse range of qudit Hamiltonians.
In particular, in conjunction with linear programming methods for robust engineering of qubit \cite{Choi2020} and qudit Hamiltonians \cite{choi_dynamical_2017, Zhou2024}, we may extend to the simulation of models with non-diagonal interaction terms, and further, it may be possible to engineer the magnetic field terms required for the AHM at higher spatial dimensions with proposed analog-digital methods to simulate multi-body interactions \cite{petiziol_quantum_2021}. Scaling to more spatial dimensions would be a valuable step towards producing a simulation that could eventually outperform classical MPS-based tensor network methods in the rapidly growing entanglement entropy regimes \cite{Banuls:2019bmf,Magnifico:2024eiy,Shinjo:2024vci}.}

Finally, rather than scaling up the simulation spatially, we may consider increasing the number of states per site. Higher-level quantum operations have been demonstrated on transmons similar to the ones used in this study \cite{nguyen_empowering_2024, champion_analog_2025}, leaving open the possibility of performing the AHM simulation with a higher level of truncation of the field operators $l\geq 2$ with both digital and analog-digital methods. \revision{We expect this to further increase the gains provided by performing a qudit-based simulation rather than a qubit-based one. At the most basic level, the number of transmons required to perform a $d$-level truncation would decrease by a factor of $\log_2{d}$ if it were done with qudits rather than qubits, and the concomitant reduction in entangling gates would provide further advantages \cite{wang_qudits_2020}.}
\section{Conclusion}\label{sec:conclusions}
In summary, we have realized qutrit-based digital and analog-digital hybrid quantum simulations of the Abelian Higgs model on transmon devices and we have demonstrated that they provide a scalable trajectory for simulating lattice gauge theories and spin-1 quantum systems. Our results show that both protocols can extract physical observables relevant to confinement within the coherence limit of present-day transmon processors.
\revision{Both methods show significant reductions in the number of transmons and entangling operations required, when compared to a theoretical best-case qubit-based implementation, and by demonstrating their feasibility despite the additional challenges associated with controlling more levels and more complicated interaction Hamiltonians, this study lends support to the idea of using qudits for the quantum simulation of naturally $d$-level systems.}

While the nature of this study does not admit a quantitative comparison between the two methods, tools developed in this work can be used for a quantitative resource analysis of the two protocols for different lattice field theory models in the future. \revision{Both the analog-digital and digital protocols appear scalable based on our numerical simulations. Challenges to scaling up the digital qutrit simulation are similar to those involved in scaling digital qubit processors, and many of the same techniques can be applied \cite{morvan_optimizing_2022}.} We find that scaling up the analog-digital protocol practically introduces more constraints, but has the potential to reduce the total number of pulses required for the simulation. An exciting next step would be to
investigate string breaking and confinement phenomena in more spatial dimensions, as well as approaching a higher-order approximation of the base theory with higher-dimensional qudits.

\begin{acknowledgments}
We would like to thank Gabriel T. Landi for helpful discussions on the Lindblad master equation.
MA and AFK were supported by the U.S. Department of Energy, Advanced Scientific Computing Research, under contract number DE-SC0025430.
YM and ZO were supported in part by the Dept. of Energy
under Award Number DE-SC0019139.
RWP, AIM, and MSB were supported by the U.S. Department of Energy, award No. DE-SC0024714. NG was supported by the U.S. Department of Energy, Office of Science, Office of Advanced Scientific Computing Research Quantum Testbed Program under contract DE-AC02-05CH11231. Devices used in the analog-digital hybrid experiment were fabricated and provided by the Superconducting Qubits at Lincoln Laboratory (SQUILL) Foundry at MIT Lincoln Laboratory, with funding from the Laboratory for Physical Sciences (LPS) Qubit Collaboratory. The traveling-wave parametric amplifier (TWPA) used in the analog-digital hybrid experiment was provided by IARPA and Lincoln Labs.
\end{acknowledgments}

\revision{\section*{Data Availability}
The data and code that support the findings of this article are openly available \cite{asaduzzaman_2026_22086663}.}

\appendix

\section{Average Hamiltonian Theory \label{app_AHT}}

For complete reviews of Average Hamiltonian Theory (AHT), see, e.g., \cite{haeberlen_coherent_1968, brinkmann2016introduction}. Here, for convenience, we briefly summarize the key ideas and present the results that are relevant to the development of our simulation protocols.

A unitary operator $U(t, t_0)$ can be written as the matrix exponential of some Hermitian operator $\Theta_{t,t_0}$: $U(t, t_0) = \exp\left(i\Theta_{t,t_0}\right)$ \cite{magnus_exponential_1954,blanes_magnus_2009}. This implies that, for an arbitrary time-dependent Hamiltonian $H(t)$ that generates $U(t_f, t_0)$ over the fixed interval $[t_0, t_f]$ by
\begin{equation}
U(t_f, t_0) = \hat{\mathcal{T}}\exp\left(-i\intf{t_0}{t_f}{H(t')}{t'}\right),
\end{equation}
where $\hat{\mathcal{T}}\exp(\cdot)$ is the time-ordered exponential, we may define some time-independent Hamiltonian $\bar{H}_{t_f,t_0}$ such that
\begin{equation}
U(t_f, t_0) = \exp\left(-iT \bar{H}_{t_f,t_0}\right),
\end{equation}
where $T = t_f - t_0$. Assuming conditions of convergence apply, the Hamiltonian $\bar{H}_{t_f,t_0}$ may be expressed with the Magnus expansion \cite{magnus_exponential_1954,blanes_magnus_2009}
\begin{equation}
\bar{H} = \sum_{n=1}^\infty \bar{H}^{(n)},
\end{equation}
where the first three terms are
\begin{equation}
\bar{H}^{(1)} = \frac{1}{T}\intf{\!t_1=t_0}{t_f}{H(t_1)}{t_1},\label{eq:magnus1}
\end{equation}
\begin{equation}
\bar{H}^{(2)} = \frac{1}{2iT}\intf{\!t_1=t_0}{t_f}{\intf{\!t_2=t_0}{t_1}{[H(t_1),H(t_2)]}{t_2}}{t_1},\label{eq:magnus2}
\end{equation}
\begin{equation}
\begin{aligned}
\bar{H}^{(3)} = &\frac{-1}{6T}\intf{\!t_1=t_0}{t_f}{\intf{\!t_2=t_0}{t_1}{\intf{\!t_3=t_0}{t_2}{\Big(\big[H(t_1),\left[H(t_2),H(t_3)\right]\big]\\
&+\big[\left[H(t_1),H(t_2)\right],H(t_3)\big]\Big)}{t_3}}{t_2}}{t_1}.
\end{aligned}\label{eq:magnus3}
\end{equation}
Truncating the series at a finite order yields an \emph{effective average Hamiltonian} $\bar{H}_\mathrm{eff} = \sum_{n=1}^m \bar{H}^{(n)}\approx \bar{H}.$

Each successive order of the average Hamiltonian involves increasingly nested commutators of $H(t)$ with itself at different times. Hence, if the Hamiltonian commutes with itself at all times, $$[H(t_1),H(t_2)] = 0 \;\forall t_1,t_2 \in [t_0, t_f],$$ then $\bar{H}^{(n)}=0$ for $n\geq 2$ and the expansion is exact at the first order.

Another property of the expansion that aids in the development of our simulation protocol is the fact that a symmetric Hamiltonian, where $H(t) = H(t_f-(t-t_0))$ for all $t\in [t_0, t_f]$, will yield exactly zero for all even-ordered terms in the expansion, that is, $\bar{H} = \bar{H}^{(1)}+\bar{H}^{(3)} + \cdots$ \cite{brinkmann2016introduction}.

Lastly, given a sequence of time-independent Hamiltonians $\{H_1, H_2, \dots, H_N\}$, or, as in the case of the analog-digital protocol, a Hamiltonian where the evolution is driven in pulsed segments that can themselves be approximated by time-independent Hamiltonians with AHT, the integrals above can be replaced by sums:
\begin{equation}
\bar{H}^{(1)}_\mathrm{disc} = \frac{1}{N}\sum_{i=1}^N H_i,\label{eq:aht_discrete1}
\end{equation}
\begin{equation}
\bar{H}^{(2)}_\mathrm{disc} = \frac{1}{2iN}\sum_{i=2}^N\sum_{j=1}^{i-1} [H_i,H_j]T,\label{eq:aht_discrete2}
\end{equation}
where we have omitted the third order term for simplicity, and where we have assumed that all Hamiltonians in the sequence are active for the same duration $T$. The ``disc'' suffix is only to distinguish this discrete average Hamiltonian from the integral definitions above.\\

\section{Analog Detuned Drive Rotating Frame Transformation and Phase Ramping Derivation}\label{app:rotating_frame}
In the analog protocol, the driven transmon Hamiltonian approximates the Abelian Higgs Hamiltonian in the frame of the drive, after accounting for the fact that the drive frequency is time-dependent. This appendix expands on the transformation into the drive frame and derives the phase ramping required to produce an instantaneous detuning with a particular envelope. We start with the lab frame Hamiltonian of the driven transmon, truncated to three levels:
\begin{equation}
\begin{aligned}
    H_k^\mathrm{LF}(t) &= H^0_k + H^\mathrm{drive}_k(t)\\
    &= \omega_{01} \ketbra{1}{1} +(\omega_{01} +\omega_{12}) \ketbra{2}{2} \\
    &+V(t) \left(i\ketbra{1}{0} + i\lambda \ketbra{2}{1}  + \mathrm{h.c.}\right).
    \label{eq:single_transmon_lab_frame_2}
\end{aligned}
\end{equation}
For time-independent detunings, $V(t)$ may be defined as in the main text, $V(t) = \sum_{i=1}^{2}\Omega_i(t)\cos(\omega_{i}^\mathrm{d}t + \phi_i) = \sum_{i=1}^{2}\Omega_i(t)\cos\left(\left(\omega_{i-1,i}-\Delta_i\right)t + \phi_i\right)$. Here we use a more general expression,
\begin{equation}
    V(t) = \sum_{i=1}^{2}\Omega_i(t)\cos(\omega_{i-1,i}t + \delta\phi_i(t)+\phi_i),
\end{equation}
where the (now potentially time-dependent) detuning has been absorbed into a time-dependent phase. The rotating frame transformation is defined in a similar manner to the time-independent case:
\begin{equation}
H^\mathrm{DF} = U_{\mathrm{D}}^\dagger H^\mathrm{LF} U_{\mathrm{D}}-i U_{\mathrm{D}}^\dagger \partial_t U_{\mathrm{D}},
\end{equation}
where now
\begin{equation}
U_{\mathrm{D}}(t)=\begin{pmatrix}
    1 & 0 & 0\\
    0 & e^{-i(\omega_{01} t+\delta\phi_1(t))} & 0\\
    0 & 0 & e^{-i ((\omega_{01}+ \omega_{12}) t+\delta\phi_1(t)+\delta\phi_2(t)}
\end{pmatrix}.
\end{equation}
This produces the following rotated Hamiltonian, without taking the RWA:
\begin{widetext}
\begin{align*}
\begin{aligned}
    H_k^\mathrm{DF}(t) &= -\delta\dot\phi_1 \ketbra{1}{1} -(\delta\dot\phi_1+\delta\dot\phi_2) \ketbra{2}{2} \\
    &\begin{aligned}   +\frac{i}{2}\Big(\Omega_1e^{-i\phi_1}
                       &+\Omega_1 e^{2i(\omega_{01} t +\delta\phi_1(t))+i\phi_1}\\
                       &+\Omega_2 e^{i(\omega_{01} t+\delta\phi_1(t)+\omega_{12} t+\delta\phi_2(t))+i\phi_2}\\
                       &+\Omega_2 e^{i(\omega_{01} t+\delta\phi_1(t)-\omega_{12} t-\delta\phi_2(t))-i\phi_2}\Big)\ketbra{1}{0}\end{aligned}\\
    &\begin{aligned}   +\frac{i\lambda}{2}\Big(\Omega_2e^{-i\phi_2}
                              &+\Omega_2 e^{2i(\omega_{12} t+\delta\phi_2(t))+i\phi_2}\\
                              &+\Omega_1 e^{i(\omega_{12} t+\delta\phi_2(t)+\omega_{01} t+\delta\phi_1(t))+i\phi_1}\\
                              &+\Omega_1 e^{i(\omega_{12} t+\delta\phi_2(t)-\omega_{01} t-\delta\phi_1(t))-i\phi_1}\Big) \ketbra{2}{1}  + \mathrm{h.c.}\end{aligned}.
    \label{eq:single_transmon_lab_frame_3}
\end{aligned}
\end{align*}
\end{widetext}

The oscillating terms involving the sums of frequencies $\omega_{01}$ and $\omega_{12}$ may be dropped safely under the same conditions as the standard RWA, assuming $\Omega \ll \omega_{01} + \omega_{12} \approx 2\omega_{01}$. The terms involving the difference of the two frequencies, however, introduce a stricter condition for the RWA: $\Omega \ll |\omega_{01}-\omega_{12}|$ (where we have already required that the oscillation due to the $\delta\phi(t)$ terms is small because the detunings must be much smaller than the magnitude of the anharmonicity). In the experiments presented here, we have $\Omega / |\omega_{01}-\omega_{12}|\sim \num{4e-3}$ in the single-qutrit experiment and $\Omega / |\omega_{01}-\omega_{12}|\sim \num{4e-4}$ in the two-qutrit experiment, which both meet the stricter RWA requirement. When the RWA conditions are met, all oscillating terms may be dropped and the rotating frame Hamiltonian may be approximated as
\begin{equation}
\begin{aligned}
    H_k^\mathrm{DF}(t) &\approx -\delta\dot\phi_1 \ketbra{1}{1} \\
    &-(\delta\dot\phi_1+
    \delta\dot\phi_2) \ketbra{2}{2} \\
    &-\frac{i}{2}\Omega_1\ketbra{0}{1}-\frac{i\lambda}{2}\Omega_2\ketbra{1}{2}  + \mathrm{h.c.}.
    \label{eq:single_transmon_lab_frame_4}
\end{aligned}
\end{equation}

The derivatives of the phase accumulation functions can be seen to play the same role as the detunings in the time-independent case, Eq.~\eqref{eq:H_rwa}, so we may define $\Delta_i(t)=\delta\dot\phi_i(t)$ to obtain,
\begin{equation}
\begin{aligned}
    H_k^\mathrm{DF}(t) &\approx -\Delta_1(t) \ketbra{1}{1} -\left(\Delta_1(t)+\Delta_2(t)\right) \ketbra{2}{2} \\
    &+\frac{i}{2}\Omega_1\ketbra{0}{1}+\frac{i\lambda}{2}\Omega_2\ketbra{1}{2}  + \mathrm{h.c.}.
    \label{eq:single_transmon_lab_frame_5}
\end{aligned}
\end{equation}
This implies that, given a desired $\Delta(t)$, we must ramp the phase on the drives according to $\delta\phi(t)=\intf{0}{t}{\Delta(t')}{t'}$ \cite{gambetta_analytic_2011}. Further, for a sequence of pulses, we find that to maintain a consistent phase reference with the detuned frame we must ensure that the phase accumulations $\delta\phi(t)$ are maintained from the end of one pulse to the start of the next.

The envelope function used for both the analog pulses and the $\pi_{02}$ gates was a flat-topped pulse with a cosine-shaped rise and fall, given by
\begin{equation}
\begin{aligned}
A_T(t)=
    \begin{cases}
        a\sin^2{\left(\frac{\pi t}{2rT}\right)}, &  0\leq t < rT\\
        a, & rT\leq t < (1-r)T\\
        a\sin^2{\left(\frac{\pi (T-t)}{2rT}\right)}, &  (1-r)T\leq t \leq T,
    \end{cases}
\end{aligned}\label{eq:analog_envelope}
\end{equation}
where $r$ is the fraction of the pulse each ramp occupies and $a$ is the normalization factor $1/(1-r)$, which results in a total area of $\intf{0}{T}{A_T(t)}{t}=T$. For the $\pi_{02}$ gates, we also incorporated a second-order DRAG correction \cite{li_universal_2025, gambetta_analytic_2011} to reduce the effects of off-resonant excitations.

\section{Dispersive Interaction Rate Calculations}\label{app:interaction_rates}
The effective dispersive interaction between two capacitively coupled transmon qutrits may be approximately derived from the time independent bare system Hamiltonian,
\begin{equation}
    H_{\mathrm{sys}} = H^0_A + H^0_B + H_{AB},
\end{equation}
where $H^0_A$ and $H^0_B$ are the bare Hamiltonians of transmons $A$ and $B$, respectively, and $H_{AB} = J \hat{n}_A \hat{n}_B \approx J (\hat{a}_A \hat{a}_B^\dagger + \hat{a}_A^\dagger \hat{a}_B)$ represents the bare capacitive coupling, where we have already taken the rotating wave approximation, with $\hat{a}_k^\dagger=\sum_{i=1}^{d} \mu_{k,i}\ketbra{i}{i-1}$, the raising operator for the energy eigenstates of transmon $k\in\{A,B\}$ with dimensionless charge matrix elements $\mu_{k,i}\approx\sqrt{i}$ \cite{Koch:2007hay}. In practice, we do not use the $\sqrt{i}$ approximation, and instead use values calculated from numerical diagonalization of the transmon Hamiltonian \cite{groszkowski_scqubits_2021,chitta_computer-aided_2022}.

We may use second-order time-independent perturbation theory \cite{messiah_quantum_1962} to approximately diagonalize the two-transmon Hamiltonian $\tilde{H}_\mathrm{sys}$ and arrive at the effective energy level shifts in the absence of any drive.
Note that we must take both transmon Hamiltonians to at least eigenstate $\ket{3}$ (four level approximation) for accurate estimates of the shifts on the first three levels.
The diagonalized Hamiltonian may be expressed generally as
\begin{equation}
    H_{\mathrm{sys}} \approx H^0_A + H^0_B + \sum_{i,j=0}^{d-1} \varepsilon_{i,j}\ketbra{ij}{ij}.
\end{equation}

After absorbing the shifts that correspond to redefining the transition frequencies of each of the two transmons into $H^0_A\rightarrow {H^{0'}_A}$ and $H^0_B\rightarrow {H^{0'}_B}$, we arrive at
\begin{equation}
    H_{\mathrm{sys}} \approx {H^{0'}_A} + {H^{0'}_B} + H_\mathrm{ck},
\end{equation}
with the cross-Kerr interaction Hamiltonian,
\begin{equation}
\begin{aligned}
    H_\mathrm{ck} &= \alpha^0_{11}\ketbra{11}{11}+\alpha^0_{12}\ketbra{12}{12}\\
    &+\alpha^0_{21}\ketbra{21}{21}+\alpha^0_{22}\ketbra{22}{22},\label{eq:interaction_ck}
\end{aligned}
\end{equation}
with rates given by the perturbation theory:
\begin{widetext}
\begin{align}
    \alpha_{11}^0&= J^2 \left(\frac{\mu_{A,1}^2 \mu_{B,2}^2}{\omega_1^A-\omega_2^B}-\frac{\mu_{A,2}^2 \mu_{B,1}^2}{\omega_2^A-\omega_1^B}\right)\\
    \alpha_{12}^0&= -J^2 \left(\frac{\mu_{A,2}^2 \mu_{B,2}^2}{\omega_2^A-\omega_2^B}+\mu_{A,1}^2 \left(\frac{\mu_{B,1}^2}{\omega_1^A-\omega_1^B}-\frac{\mu_{B,2}^2}{\omega_1^A-\omega_2^B}-\frac{\mu_{B,3}^2}{\omega_1^A-\omega_3^B}\right)\right)\\
    \alpha_{21}^0&= J^2 \left(\frac{\mu_{A,2}^2\mu_{B,2}^2}{\omega_2^A-\omega_2^B}+\mu_{B,1}^2\left(\frac{\mu_{A,1}^2}{\omega_1^A-\omega_1^B}-\frac{\mu_{A,2}^2}{\omega_2^A-\omega_1^B}-\frac{\mu_{A,3}^2}{\omega_3^A-\omega_1^B}\right)\right)\\
    \alpha_{22}^0&= -J^2 \left(\mu_{A,2}^2 \left(\frac{\mu_{B,1}^2}{\omega_2^A-\omega_1^B}-\frac{\mu_{B,3}^2}{\omega_2^A-\omega_3^B}\right)-\mu_{B,2}^2 \left(\frac{\mu_{A,1}^2}{\omega_1^A-\omega_2^B}-\frac{\mu_{A,3}^2}{\omega_3^A-\omega_2^B}\right)\right).
\end{align}
\end{widetext}
For the sake of visual simplicity, in this section we represent the transition frequency between levels $i-1$ and $i$ as $\omega_i\equiv\omega_{i-1,i}$.

To modify the interaction rates, we use Stark drives at frequency $\omega_S$ applied to both qutrits. This method has been used to produce controlled-Z gates in both superconducting qubit \cite{mitchell_hardware-efficient_2021, wei_hamiltonian_2022} and qutrit devices \cite{goss_high-fidelity_2022}. For convenience, we review the perturbation analysis of the method here.

We start with a two-transmon Hamiltonian in the lab frame with Stark drives on both transmons,
\begin{equation}
    \tilde{H}_\mathrm{sys}^\mathrm{LF} = H^0_A + H^\mathrm{Stark}_A(t) + H^0_B + H^\mathrm{Stark}_B(t) + H_{AB},
\end{equation}
where the Stark drive Hamiltonian for each transmon, assuming the rotating wave approximation from the outset, is given by
\begin{equation}
\begin{aligned}
    H^\mathrm{Stark}_k(t) &= i\Omega_{S,k}^0 \cos{(\omega_{S}t+\phi_k)}(\hat{a}_k^\dagger - \hat{a}_k)\\
    &\approx  i\frac{\Omega_{S,k}^0}{2} \left(e^{-i(\omega_{S}t+\phi_k)}\hat{a}_k^\dagger - e^{i(\omega_{S}t+\phi_k)}\hat{a}_k\right).
\end{aligned}
\end{equation}
We may obtain a time-independent RWA Hamiltonian by rotating into the frame of the drive on both transmons with $U_{\mathrm{S}}=U_{\mathrm{S},A}\sotimes U_{\mathrm{S},B}$ and $U_{\mathrm{S},k}=\sum_{n=0}^d e^{in\omega_S t}\ketbra{n}{n}$, arriving at a time-independent Hamiltonian in the frame of the Stark drive,

\begin{equation}
\begin{aligned}
    \tilde{H}^\mathrm{SF}_\mathrm{sys} &\approx \left(\sum_{i=1}^{d} \epsilon_i^A\ketbra{i}{i}\right)\sotimes I + I\sotimes\left(\sum_{j=1}^{d} \epsilon_j^B \ketbra{j}{j}\right) \\
    &+ \left(i \sum_{i=1}^{d}\mu_{A,i}\frac{\Omega_{S,A}^0 e^{-i\phi_A}}{2}\ketbra{i}{i-1}\right)\sotimes I \\
    &+ I\sotimes \left(i \sum_{j=1}^{d}\mu_{B,j}\frac{\Omega_{S,B}^0  e^{-i\phi_B}}{2}\ketbra{j}{j-1} \right)\\
    &+ J \sum_{i,j=1}^{d} \mu_{A,i}\mu_{B,j}\ketbra{i}{i-1}\sotimes\ketbra{j-1}{j} + \mathrm{h.c.},
\end{aligned}
\end{equation}
where we define $\epsilon^k_j \equiv \sum_{j'=1}^j (\omega_{j'}^k-\omega_S)=\sum_{j'=1}^j \Delta_{j'}^k$ and $\Delta_i^k\equiv\omega^k_{i-1,i}-\omega_S$.
That is, the effective energy terms in the rotated frame are the cumulative sums of the detunings of each successive transition from the Stark drive frequency.

The Hamiltonian may now be perturbatively diagonalized to third order, and then rotated back to the lab frame with the inverse transformation $U_\mathrm{S}^\dagger$, resulting in an effectively diagonal Hamiltonian with modified interaction rates and Stark-shifted transition frequencies on both transmons,
\begin{equation}
    \tilde{H}_{\mathrm{sys}}^\mathrm{LF} \approx {\tilde{H}^{0'}_A} + {\tilde{H}^{0'}_B} + \tilde{H}_\mathrm{ck}.
\end{equation}
The coefficients in the modified interaction Hamiltonian $\tilde{H}_\mathrm{ck}$ are
\begin{widetext}
\begin{align}
    \tilde{\alpha}_{11} &= J \Omega_{S,A}^0 \Omega_{S,B}^0 \cos(\delta\phi) \Biggl\{\frac{\mu_{A,1}^2}{\Delta_1^A} \left(\frac{2 \mu_{B,1}^2}{\Delta_1^B}-\frac{\mu_{B,2}^2}{\Delta_2^B}\right)-\frac{\mu_{A,2}^2}{2\Delta_2^A} \left(\frac{2\mu_{B,1}^2}{\Delta_1^B}-\frac{\mu_{B,2}^2}{\Delta_2^B}\right)\Biggr\}+\alpha_{11}^0\label{eq:stark_alpha_11}\\
    \tilde{\alpha}_{12} &= \begin{aligned}[t]
    J \Omega_{S,A}^0 \Omega_{S,B}^0 \cos(\delta\phi) \Biggl\{\frac{\mu_{A,1}^2}{\Delta_1^A} \left(\frac{\mu_{B,1}^2}{\Delta_1^B}+\frac{\mu_{B,2}^2}{\Delta_2^B}-\frac{\mu_{B,3}^2}{\Delta_3^B}\right)-\frac{\mu_{A,2}^2}{2\Delta_2^A} \left(\frac{\mu_{B,1}^2}{\Delta_1^B}+\frac{\mu_{B,2}^2}{\Delta_2^B}-\frac{\mu_{B,3}^2}{\Delta_3^B}\right)\Biggr\}+\alpha_{12}^0
    \end{aligned}\label{eq:stark_alpha_12}\\
    \tilde{\alpha}_{21} &= \begin{aligned}[t]
    J \Omega_{S,A}^0 \Omega_{S,B}^0 \cos(\delta\phi) \Biggl\{ \frac{\mu_{B,1}^2}{\Delta_1^B}\left(\frac{\mu_{A,1}^2}{\Delta_1^A}+\frac{\mu_{A,2}^2}{\Delta_2^A}-\frac{\mu_{A,3}^2}{\Delta_3^A}\right)-\frac{\mu_{B,2}^2}{2\Delta_2^B}\left(\frac{\mu_{A,1}^2}{\Delta_1^A } +\frac{\mu_{A,2}^2}{\Delta_2^A} -\frac{\mu_{A,3}^2}{\Delta_3^A}\right)\Biggr\}+\alpha_{21}^0
    \end{aligned}\label{eq:stark_alpha_21}\\
    \tilde{\alpha}_{22} &= \begin{aligned}[t]
    J \Omega_{S,A}^0 \Omega_{S,B}^0 \cos(\delta\phi) \Biggl\{\frac{\mu_{A,1}^2}{2 \Delta_1^A} \left(\frac{\mu_{B,1}^2}{\Delta_1^B}+\frac{\mu_{B,2}^2}{\Delta_2^B}-\frac{\mu_{B,3}^2}{\Delta_3^B}\right) &+\frac{\mu_{A,2}^2}{2 \Delta_2^A} \left(\frac{\mu_{B,1}^2}{\Delta_1^B}+\frac{\mu_{B,2}^2}{\Delta_2^B}-\frac{\mu_{B,3}^2}{\Delta_3^B}\right)\\
     &- \frac{\mu_{A,3}^2}{2 \Delta_3^A} \left(\frac{\mu_{B,1}^2}{\Delta_1^B}+\frac{\mu_{B,2}^2}{\Delta_2^B}-\frac{\mu_{B,3}^2}{\Delta_3^B}\right)\Biggr\}+\alpha_{22}^0\label{eq:stark_alpha_22},
    \end{aligned}
\end{align}
\end{widetext}
where $\delta\phi$ is the phase difference between the Stark drives, $\delta\phi=\phi_A-\phi_B$. The local frequency shifts in $\tilde{H}^{0'}_k$ are the standard AC-Stark shifts \cite{cohen-tannoudji_atom-photon_2004} of a multilevel ladder-type system subjected to an off-resonant drive at $\omega_S$, that is, $\tilde{\omega}_{i-1,i}=\omega_{i-1,i}+\delta^\mathrm{S}_{i-1,i}$, where
\begin{equation*}
\delta^\mathrm{S}_{i-1,i}=\frac{-1}{4}\frac{\mu_{i-1}^2\Omega_S^2}{\Delta_{i-1}}+\frac{1}{2}\frac{\mu_{i}^2\Omega_S^2}{\Delta_{i}}-\frac{1}{4}\frac{\mu_{i+1}^2\Omega_S^2}{\Delta_{i+1}},
\end{equation*}
and we take $\mu_0\equiv0$. We measure these local shifts with Ramsey experiments and compensate for them by adjusting the detunings of the on-site simulation drives.

Given the above expressions for $\tilde{\alpha}_{ij}(\Omega_{S,A}^0,\Omega_{S,B}^0,\omega_S,\delta\phi)$, we now perform a transformation of variables to simplify the engineering of the desired $L_z\sotimes L_z$ interaction Hamiltonian. For convenience, we will use an intermediate representation that better respects the structure of the target Hamiltonian. By analogy with the definition of the $\alpha_{ij}$ in terms of the energy levels of the two-transmon system, $\alpha_{ij}=(\varepsilon_{ij}-\varepsilon_{i0})-(\varepsilon_{0j}-\varepsilon_{00})$, we may define instead $\zeta_{ij}=(\varepsilon_{ij}-\varepsilon_{i1})-(\varepsilon_{1j}-\varepsilon_{11})$, which contain the same information as the $\alpha_{ij}$, but reference energy levels from the $\ket{11}$ state, rather than the $\ket{00}$ state of the two-transmon system. By substituting the expressions for $\alpha_{ij}$ into this definition, we find four nonzero $\zeta_{ij}$:
\begin{equation}
\begin{aligned}
    \zeta_{00} &= \alpha_{11}\\
    \zeta_{02} &= \alpha_{11}-\alpha_{12}\\
    \zeta_{20} &= \alpha_{11}-\alpha_{21}\\
    \zeta_{22} &= (\alpha_{22}-\alpha_{21})-(\alpha_{12}-\alpha_{11}).
\end{aligned}\label{eq:zeta_alpha_conversion}
\end{equation}
This procedure effectively redefines the transition frequencies of the individual transmons, as outlined in the main text, and results in a Hamiltonian that is equivalent to that given in Eq.~\eqref{eq:zz_coeffs},  represented in a different basis:
\begin{equation}
\begin{aligned}
    H_\mathrm{ck}' &= \zeta_{00}\ketbra{00}{00}+\zeta_{02}\ketbra{02}{02}\\
    &+\zeta_{20}\ketbra{20}{20}+\zeta_{22}\ketbra{22}{22}.\label{eq:interaction_ckzz}
\end{aligned}
\end{equation}

\begin{figure*}[!ht]
\captionsetup[subfigure]{labelformat=nocaption}
\includegraphics[width=0.9\linewidth]{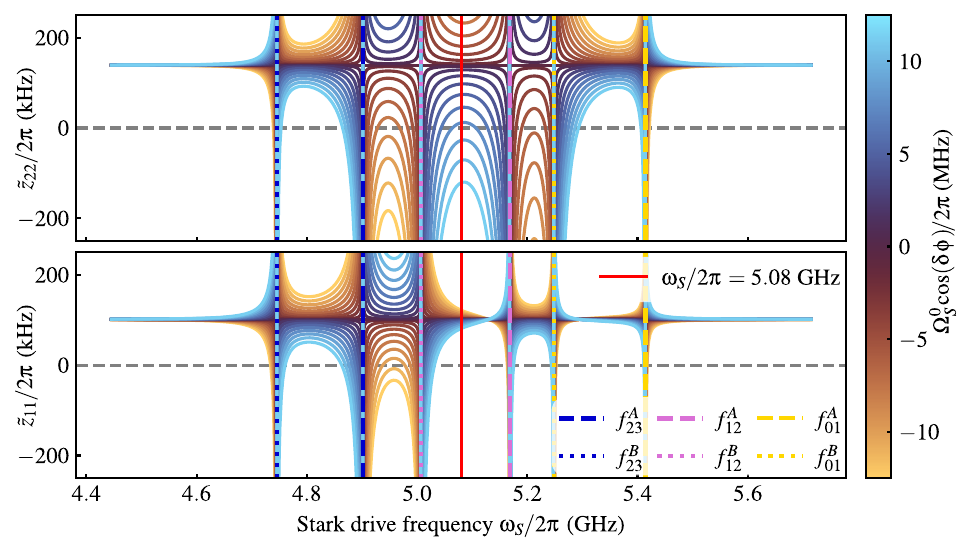}%
\caption{Third-order perturbation theory results for the driven interaction rates, based on the experimentally measured spectra of the two transmons used in the analog-digital experiment, plotted vs frequency for different Stark drive amplitudes $\Omega_S^0$ and phase differences $\delta\phi\in \{0,\pi\}$. Transition frequencies of each transmon $k\in\{A,B\}$, $f_{ij}^k=\omega_{ij}^k/2\pi$, are marked with vertical lines. Expressions for the driven rates $\tilde{z}_{11}(\Omega_{S}^0,\omega_S,\delta\phi)$ and $\tilde{z}_{22}(\Omega_{S}^0,\omega_S,\delta\phi)$ can be obtained from Eqs.~\eqref{eq:stark_alpha_11}--\eqref{eq:stark_alpha_22}. The solid red vertical line indicates the frequency of the Stark drive used in this experiment.
}\label{fig:stark_freq_theory}
\end{figure*}

From here, it is a matter of inspection to define the $z_{ij}$ coefficients in terms of the $\zeta_{ij}$ and hence the $\alpha_{ij}$. This relationship may be represented by the matrix equation,
\begin{equation}
    \begin{pmatrix}
        z_{11}\\
        z_{12}\\
        z_{21}\\
        z_{22}
    \end{pmatrix}=
    \frac{1}{4}\begin{pmatrix}
     1 & -1 & -1 & 1 \\
     1 & 1 & -1 & -1 \\
     1 & -1 & 1 & -1 \\
     1 & 1 & 1 & 1
    \end{pmatrix}\begin{pmatrix}
        \zeta_{00}\\
        \zeta_{02}\\
        \zeta_{20}\\
        \zeta_{22}
    \end{pmatrix}.
\end{equation}

To identify candidate Stark drive frequencies and amplitudes, we perform a numerical search for zero crossings with the perturbation theory expression for the driven $\tilde{z}_{22}(\Omega_{S,A}^0,\Omega_{S,B}^0,\omega_S,\delta\phi)$. We simplify the parameter space by restricting $\delta\phi$ to two values $\delta\phi\in\{0,\pi\}$, and by requiring that $\Omega_{S,A}^0 = \Omega_{S,B}^0 = \Omega_S^0$. We find that for the two transmons used here (in the straddling regime), there exist several ranges of frequencies that admit $\tilde{z}_{22}=0$ for a given drive amplitude. As can be seen in Fig.~\ref{fig:stark_freq_theory}, we see the strongest response in $\tilde{z}_{22}$ when $\omega_{12}^B<\omega_S<\omega_{12}^A$. We find that the perturbation theory results agree at least qualitatively with the experiment when not driving too close to any transition frequencies, and are therefore a useful starting point for the calibration. Experimental calibration of the amplitudes is described in Appendix~\ref{app:calibration_sequences}, see Fig.~\ref{fig:interaction_calibration}.

\section{Entangling Rate Measurement Protocols}\label{app:calibration_sequences}

\begin{figure*}[!htb]
\captionsetup[subfigure]{labelformat=nocaption}
\includegraphics[width=0.9\linewidth]{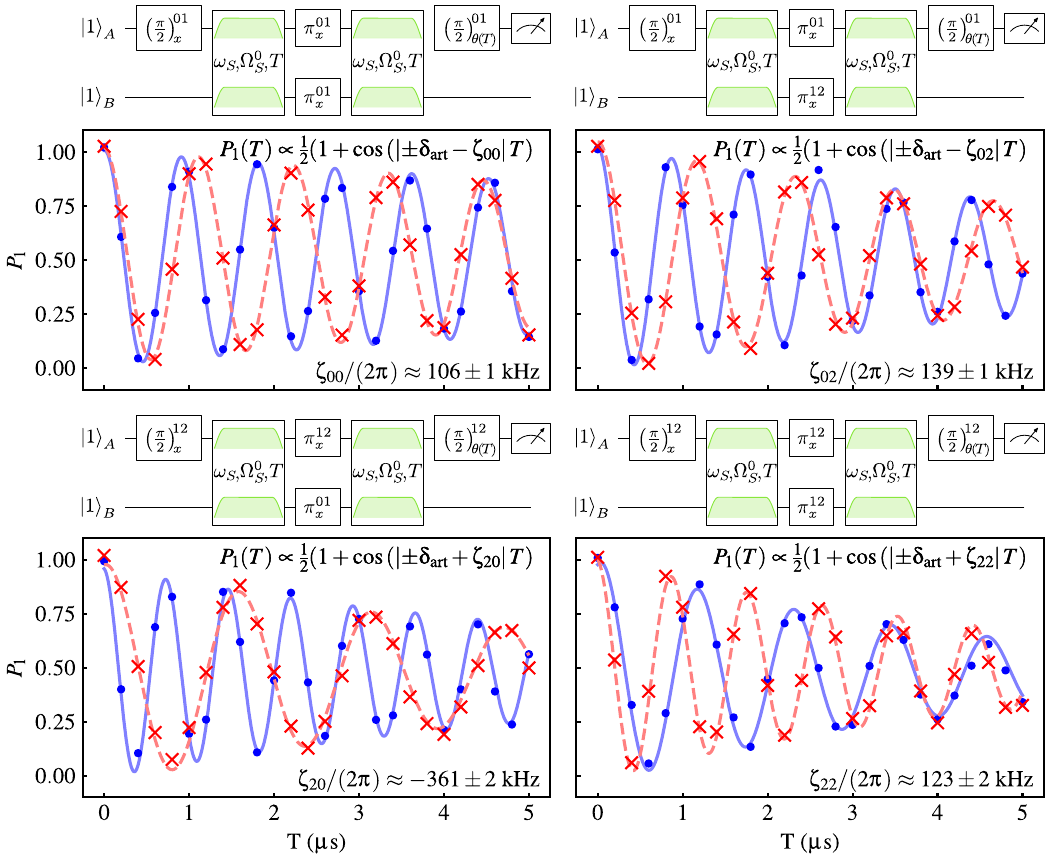}%
\caption{JAZZ sequence circuit diagrams and measurement results to estimate Stark-driven $\tilde{\zeta}_{ij}$ entangling rates. Each circuit is run twice, with $\theta(T) = \pm\delta_\mathrm{art.}T$ ($+$: blue circles, solid line fit; $-$: red crosses, dashed line fit), and the signed values of $\zeta_{ij}$ are extracted from exponentially decaying cosine fits of the two measurements. The artificial detuning was $\delta_\mathrm{art}/2\pi=\qty{1}{\MHz}$ for all measurements. These measurements were taken with a calibrated Stark drive, yielding $\tilde{z}_{22}=\frac{1}{4}(\tilde{\zeta}_{00}+\tilde{\zeta}_{02}+\tilde{\zeta}_{20}+\tilde{\zeta}_{22}) \approx \qty{-2}{\kHz}$.}
\label{fig:jazz_sequences}\label{fig:jazz_results}
\end{figure*}

To measure the coefficients of the entangling Hamiltonian experimentally, we use a series of modified JAZZ/BIRD sequences \cite{takita_experimental_2017, garbow_bilinear_1982} that can both extract the individual entangling rates and decouple from small transition frequency offsets produced by miscalibration or slow drifts.

For two qutrits, each entangling rate $\zeta_{ij}$ ($i,j\in\{0,2\}$)---defined in Eq.~\eqref{eq:interaction_ckzz} and related to the original cross-Kerr entangling rates $\alpha_{ij}$ ($i,j\in\{1,2\}$) in Eq.~\eqref{eq:zeta_alpha_conversion}---may be extracted from the Ramsey oscillations of an individual JAZZ sequence, as shown in Fig.~\ref{fig:jazz_sequences}, where the sequences for different $i,j$ only differ by the choice of subspace for the $\pi/2$ pulses and the $\pi$ pulses that echo any unconditional frequency shifts.

To unambiguously determine the sign of the JAZZ Ramsey frequencies, we add an ``artificial detuning'' $\delta_\mathrm{art}$ to the Ramsey sequence by varying the phase of the final $\pi/2$ pulse as $\theta(T) = \pm\delta_\mathrm{art}T$, where $T$ is the (varying) duration of each evolution period in the sequence \cite{gao_practical_2021}, and we perform each JAZZ Ramsey experiment twice: once with $+\delta_\mathrm{art}$, and once with $-\delta_\mathrm{art}$. Specific to this particular decoupling pulse protocol, the sign of the measured JAZZ Ramsey frequencies is reversed for $i=2$, which must be compensated for.

The frequency of oscillation in a single JAZZ measurement for a particular $\zeta_{ij}$ is then $\omega_{\mathrm{JAZZ},0j}^\pm = \left|\pm\delta_\mathrm{art}-\zeta_{0j}\right|$ for $i=0$ or $\omega_{\mathrm{JAZZ},2j}^\pm = \left|\pm\delta_\mathrm{art}+\zeta_{2j}\right|$ for $i=2$. Under the assumption that the magnitude of the artificial detuning is larger than the magnitude of the frequency to be measured, the signed entangling rate may be estimated by,
\begin{equation}
\begin{aligned}
    \zeta_{ij} = \begin{cases}
        \frac{\omega_{\mathrm{JAZZ},ij}^{-}-\omega_{\mathrm{JAZZ},ij}^{+}}{2}, & i=0\\
        \frac{\omega_{\mathrm{JAZZ},ij}^{+}-\omega_{\mathrm{JAZZ},ij}^{-}}{2}, &i=2
    \end{cases}
\end{aligned}
\end{equation}
Individual JAZZ Ramsey measurement results for the driven rates $\tilde{\zeta}_{ij}$ are shown in Fig.~\ref{fig:jazz_results} below each corresponding circuit diagram.

\begin{figure*}[!htb]
\captionsetup[subfigure]{labelformat=nocaption}
\includegraphics[width=0.9\linewidth]{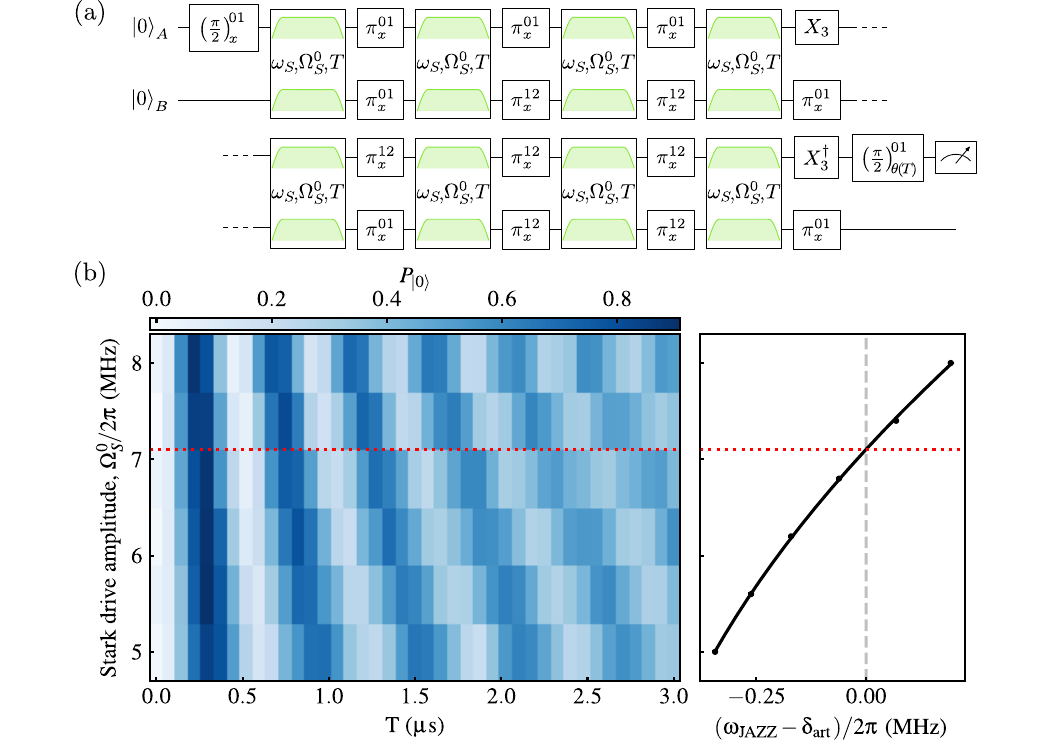}%
\begin{subfigure}{0\linewidth}
\caption{}\label{fig:zeta_sum_sequence}
\end{subfigure}%
\begin{subfigure}{0\linewidth}
\caption{}\label{fig:interaction_calibration}
\end{subfigure}%
\caption{Circuit diagram and calibration measurement for the Stark drive amplitude. (a) ``Total JAZZ'' sequence, designed to produce oscillations in $T$ at a frequency corresponding to $\zeta_{00}+\zeta_{02}+\zeta_{20}+\zeta_{22} = 4z_{22}$. The circuit is split into two lines for clarity. The ``shift'' gate $X_d=\sum_n^{d-1} \ket{n+1 \pmod{d}}\bra{n}$ (see, e.g. \cite{schwinger_unitary_1960,gottesman_fault-tolerant_1999}) may be performed on a qutrit with two $\pi$ pulses: $X_3=\pi_x^{01}\pi_x^{12}$. (b) Stark drive amplitude calibration measurement with the ``Total JAZZ'' sequence. Given a particular $\omega_S$, we perform the circuit in \subref{fig:zeta_sum_sequence} for varying Stark drive amplitudes $\Omega_S^0$ and varying drive durations $T$. At each value of $\Omega_S^0$, the Ramsey oscillations in the left panel are fit with an exponentially decaying cosine function, and the resulting oscillation frequencies, with the artificial detuning of $\delta_\mathrm{art}=\qty{3}{\MHz}$ subtracted, are shown in the right panel. The frequencies are then fit with a quadratic as a function of $\Omega_S^0$, and the zero crossing point is solved for analytically, providing the calibrated Stark drive amplitude.}
\end{figure*}

In addition, the more complicated sequence given in Fig.~\ref{fig:zeta_sum_sequence}, with eight periods of evolution, may be shown to produce oscillations at the sum of the $\zeta_{ij}$ interaction rates, providing an estimate of $z_{22}=\frac{1}{4}(\zeta_{00}+\zeta_{02}+\zeta_{20}+\zeta_{22})$ in a single experiment. This allows for efficient calibration of the Stark drive amplitudes required to produce the condition $z_{22}=0$, as shown in Fig.~\ref{fig:interaction_calibration}.

Finally, to calibrate the local frequency shifts produced by the Stark drives on the two transmons, we perform simple Ramsey experiments on each transition of each transmon with the Stark drive active during the wait period. These measurements are performed while the other transmon is in the $\ket{1}$ state, to directly provide the frequencies required by the frame shift of Eq.~\eqref{eq:zz_coeffs} and also described in Appendix~\ref{app:interaction_rates}.

\section{Analog-Digital Simulation Coherence Times and Heuristic Average Decay Model}\label{app:analog_coherence_time_model}

\begin{table}[h!]
\centering
\begin{tabular}{l|ccc}
\hline
 & QA & & QB \\
\hline
$\omega_{01}/2\pi$ (GHz) & 5.4146 & & 5.2484 \\
$\omega_{12}/2\pi$ (GHz) & 5.1687 & & 5.0075 \\
$\omega_{23}/2\pi$ (GHz) & 4.918 & & 4.758 \\
$T_1^{01}$ ($\mu$s) & 30(3) & & 26(7) \\
$T_{2R}^{01}$ ($\mu$s) & 48(8) & & 36(14) \\
$T_1^{12}$ ($\mu$s) & 15(4) & & 17(5) \\
$T_{2R}^{12}$ ($\mu$s) & 20(10) & & 24(9) \\
\hline
$\omega_S$ (GHz) & 5.08 & & 5.08 \\
$\Omega_S^0$ (MHz) & 7.1 & & 7.1\\
$J$ (MHz) & & 2 & \\
\hline
\end{tabular}
\caption{Relevant parameters of the two transmons in the Rochester-based device used in the analog-digital experiment.}\label{tab:analog_coherence}
\end{table}

The coherence times and other relevant parameters of the two transmons used in the analog-digital experiment are given in Table~\ref{tab:analog_coherence}.

In the single qutrit experiment, we assume Markovian noise processes and model decoherence with the Lindblad master equation \cite{carmichael_open_1993}
\begin{equation}
\begin{aligned}
    \dot\rho &= -i[H,\rho]+\sum_\alpha \Gamma_\alpha \left(L_\alpha\rho L_\alpha^\dagger-\frac{1}{2}L_\alpha^\dagger L_\alpha\rho-\frac{1}{2}\rho L_\alpha^\dagger L_\alpha\right) \\
    &\equiv \mathcal{L}(\rho),\label{eq:lindblad_equation}
\end{aligned}
\end{equation}
where the Hamiltonian is taken to be the single-qutrit Hamiltonian in Eq.~\eqref{eq:single_qutrit_model}, $\mathcal{L}(\cdot)$ is the Liouville superoperator, and the transmon's Lindblad jump operators~\cite{li_decoherence_2011, li_pure_2012}
\begin{equation}
\begin{aligned}
&L_{10} = \ketbra{0}{1}, \quad &L_{11} = \ketbra{1}{1},\\
&L_{21} = \ketbra{1}{2}, \quad &L_{22} = \ketbra{2}{2},\label{eq:single_qutrit_jump_operators}
\end{aligned}
\end{equation}
describe the decay of the excited states ($L_{10}, L_{21}$) and pure dephasing ($L_{11}, L_{22}$).
The corresponding rates $\Gamma_{ij}$ are calculated from the coherence times reported in Table~\ref{tab:analog_coherence} as
\begin{equation}
\begin{aligned}
&\Gamma_{10} = (T_1^{01})^{-1}, \quad &\Gamma_{11} = (T_2^{01})^{-1}-\frac{1}{2}(T_1^{01})^{-1},\\
&\Gamma_{21} = (T_1^{12})^{-1}, \quad &\Gamma_{22} = (T_2^{12})^{-1}-\frac{1}{2}(T_1^{12})^{-1}.
\end{aligned}
\end{equation}

In the two-qutrit experiment, we alternate long periods of evolution ($T_E=\qty{800}{\ns}$) with relatively brief unitary operations ($T_g=\qty{60}{\ns}$) to approximate the time-independent Hamiltonian $H_\mathrm{AHM}$. To account for decoherence with a time-independent master equation model, we assume a modified set of Lindblad jump operators, which we justify here.

During each evolution period between $\pi_{02}$ flips, the evolution may be formally written $\rho(t)=e^{\mathcal{L}t}\left(\rho(0)\right)$ \cite{breuer2002theory}. By preceding the evolution period with the application of a unitary $U$ and following it with the inverse $U^\dagger$---both assumed to be instantaneous---we may write the new evolution as
\begin{equation}
    \tilde\rho(t)=\mathcal{U}^\dagger\left(e^{\mathcal{L}t}\left(\mathcal{U}\left(\rho(0)\right)\right)\right),
\end{equation}
where we define the superoperator $\mathcal{U}(\cdot)=U\cdot U^\dagger$. By expanding the exponential operator, we may see that $\mathcal{U}$, $\mathcal{U}^\dagger$ may be absorbed into the exponential:
\begin{equation}
    \mathcal{U}^\dagger\left(e^{\mathcal{L}t}\left(\mathcal{U}\left(\cdot\right)\right)\right) = e^{\mathcal{U}^\dagger \mathcal{L}t\mathcal{U}}(\cdot),
\end{equation}

leading to a new Liouvillian:
\begin{widetext}
\begin{equation}
\begin{aligned}
    (\mathcal{U}^\dagger \mathcal{L}\mathcal{U})(\rho) &= -iU^\dagger[H,U\rho U^\dagger]U+\sum_\alpha \Gamma_\alpha U^\dagger\left(L_\alpha U\rho U^\dagger L_\alpha^\dagger-\frac{1}{2}L_\alpha^\dagger L_\alpha U\rho U^\dagger-\frac{1}{2}U\rho U^\dagger L_\alpha^\dagger L_\alpha\right)U \\
    &= -i[\tilde H,\rho]+\sum_\alpha \Gamma_\alpha \left(\tilde L_\alpha\rho \tilde L_\alpha^\dagger-\frac{1}{2}\tilde L_\alpha^\dagger \tilde L_\alpha\rho-\frac{1}{2}\rho \tilde L_\alpha^\dagger \tilde L_\alpha\right) \\
    &\equiv \tilde{\mathcal{L}}(\rho),
\end{aligned}
\end{equation}
\end{widetext}
where we have defined the toggling frame Hamiltonian $\tilde H = U^\dagger H U$ and jump operators $\tilde L_\alpha = U^\dagger L_\alpha U$.

Next, assuming the evolution time $T_E$ is short compared to the timescales of the evolution---set by $H$ and the $\Gamma_\alpha$---we may approximate the exponential evolution under the Lindblad equation to first order in $T_E$ as,
\begin{equation}
\begin{aligned}
    \rho(T_E)=e^{\mathcal{L}T_E}\left(\rho(0)\right) \approx (1+\mathcal{L}T_E)(\rho(0)).
\end{aligned}
\end{equation}
Then using $\rho(T_E)$ as the initial state for evolution under $\tilde{\mathcal{L}}$, we find, again to first order,
\begin{equation}
\begin{aligned}
    \rho(2T_E)&=e^{\tilde{\mathcal{L}}T_E}\left(\rho(T_E)\right) \approx (1+\tilde{\mathcal{L}}T_E)(1+\mathcal{L}T_E)(\rho(0))\\
    & = (1+\tilde{\mathcal{L}}T_E+\mathcal{L}T_E+\tilde{\mathcal{L}}\mathcal{L}T_E^2)(\rho(0))\\
    &\approx (1+\tilde{\mathcal{L}}T_E+\mathcal{L}T_E)(\rho(0))\\
    &\approx e^{(\tilde{\mathcal{L}}+\mathcal{L})T_E}\left(\rho(0)\right).
\end{aligned}
\end{equation}
The effective average Liouville operator for the evolution over two periods is then
\begin{equation}
\begin{aligned}
    &\frac{1}{2}(\tilde{\mathcal{L}}+\mathcal{L})(\rho) = -i\frac{1}{2}[H+\tilde H,\rho]\\
    &+\frac{1}{2}\sum_{\alpha} \Gamma_{\alpha} \left(L_{\alpha}\rho L_{\alpha}^\dagger-\frac{1}{2} L_{\alpha}^\dagger L_{\alpha}\rho-\frac{1}{2}\rho L_{\alpha}^\dagger L_{\alpha}\right) \\
    &+ \frac{1}{2}\sum_{\alpha} \Gamma_{\alpha} \left(\tilde L_{\alpha}\rho \tilde L_{\alpha}^\dagger-\frac{1}{2}\tilde L_{\alpha}^\dagger \tilde L_{\alpha}\rho-\frac{1}{2}\rho \tilde L_{\alpha}^\dagger \tilde L_{\alpha}\right)\\
    &\approx \mathcal{L}_\mathrm{eff}(\rho).
\end{aligned}
\end{equation}
The Hamiltonian part of the operator is the first order average Hamiltonian as discussed in the main text, and the set of jump operators $L_{\alpha}$ are defined for each qutrit as in Eq.~\eqref{eq:single_qutrit_jump_operators}.

Because the Lindblad equation assumes Markovianity, we expect this average-Lindblad model to miss potential dynamical decoupling effects due to the periodic pulses. However, the level of agreement found between numerical evolution of the model and the data indicates that such effects do not significantly contribute to the evolution here.

\section{Error Analysis in Analog-Digital Simulation}\label{app:analog_error_terms}

In this appendix, we apply average Hamiltonian theory (AHT) to analyze coherent errors in the two-qutrit analog-digital protocol. We address two distinct sources of error: (1) errors arising from the failure of the time-varying terms in the Stark-driven Hamiltonian to commute with each other
over the course of a single pulse, and (2) errors arising from higher-order terms in the dynamical decoupling sequence.

To analyze the errors within the pulse originating from pulse shaping, we start with the system Hamiltonian in the drive frame from Eq.~\eqref{eq:analog_system_hamiltonian}:
\begin{equation}
\begin{aligned}
    H_\mathrm{sys}(t) &= A(t)\left(\Delta_0 L_z^2 + \frac{\Omega_0}{\sqrt{2}}L_x\right)\sotimes I \\
    &+ I\sotimes A(t)\left(\Delta_0 L_z^2 + \frac{\Omega_0}{\sqrt{2}}L_x\right)\\
    &+ A(t)\sum_{ij=1}^2 \delta z_{ij}L_z^i\sotimes L_z^j + \sum_{ij=1}^2 z_{ij}^0 L_z^i\sotimes L_z^j,\label{eq:System_Hamiltonian}
\end{aligned}
\end{equation}
where we define $A(t)\delta z_{ij} = \tilde{z}_{ij}(t)-z_{ij}^0$ as the part of the entangling Hamiltonian that depends on the Stark drives (see Appendix~\ref{app:interaction_rates}). Throughout this section, superscripts on the $L_\alpha$ operators will indicate powers, with the site index indicated by their position relative to the $\otimes$ symbol. Because the time-dependent parts of the Hamiltonian are all constructed to vary with the same envelope $A(t)$ (see the main text), we may write
\begin{equation}
\begin{aligned}
    H_\mathrm{sys}(t) &= A(t)M_1 + \sum_{ij=1}^2 z_{ij}^0 L_z^i\sotimes L_z^j\\
    &\equiv H(t) + H_0,
\end{aligned}
\end{equation}
where $M_1$ contains the structure of the time-dependent part of Eq.~\eqref{eq:System_Hamiltonian}.
We are then interested in taking the Magnus expansion of this Hamiltonian to third order to get the effective time-independent Hamiltonian that describes the evolution over a single evolution period. See Eqs.~\eqref{eq:magnus1}--\eqref{eq:magnus3} in Appendix~\ref{app_AHT} for definitions of the first several terms in the Magnus expansion. The first order term is given in Eq.~\eqref{eq:full_analog_first_order}, and the higher order terms produce coherent errors.

We first compute the commutators for the integrands of the second and third-order terms. Using the division of our Hamiltonian into a time-dependent and a time-independent part, we can simplify our calculation of the commutators with $[H_0, H_0] = 0$ and $[H(t), H(t')] = A(t)A(t') [M_1,M_1] = 0$, thus the only contributions will come from the commutator of the time-independent and time-dependent parts and their nested commutators. With that we can write:
\begin{align*}
    [H_\mathrm{sys}(t) , H_\mathrm{sys}(t')]& = [H(t), H_0] - [H(t'), H_0] \\&\equiv \mathcal{C}_{10}(t) -\mathcal{C}_{10}(t'),
\end{align*}
where we define $\mathcal{C}_{10}(t)$ for later convenience. The subscript refers to the time-dependence of the terms in the (nested) commutator sequence, where we use $1$ to denote $H(t)$ and a $0$ denotes $H_0$. Similarly, for the third order commutators,
\begin{align*}
    &\begin{aligned}\Big[H_\mathrm{sys}(t), \left[H_\mathrm{sys}(t'), H_\mathrm{sys}(t'')\right]\Big] &= \left[H_\mathrm{sys}(t), \mathcal{C}_{10}(t')\right] \\
            &- \left[H_\mathrm{sys}(t), \mathcal{C}_{10}(t'')\right] \end{aligned}\\
    &=  \left[H(t), \mathcal{C}_{10}(t')\right] - \left[H(t), \mathcal{C}_{10}(t'')\right]\\
    &+  \left[H_0, \mathcal{C}_{10}(t')\right] - \left[H_0, \mathcal{C}_{10}(t'')\right]\\
    &\equiv \mathcal{C}_{110}(t,t') - \mathcal{C}_{110}(t,t'') + \mathcal{C}_{010}(t') - \mathcal{C}_{010}(t'')
\end{align*}
\begin{align*}
    &\Big[\left[H_\mathrm{sys}(t), H_\mathrm{sys}(t')\right],H_\mathrm{sys}(t'')\Big] \\
    &= -\Big[H_\mathrm{sys}(t''), \left[H_\mathrm{sys}(t),H_\mathrm{sys}(t')\right]\Big]\\
    &\equiv \mathcal{C}_{110}(t'',t') - \mathcal{C}_{110}(t'',t) +  \mathcal{C}_{010}(t') - \mathcal{C}_{010}(t).
\end{align*}
Finally, we need to evaluate the forms of the operators $\mathcal{C}_{10}(t)$, $\mathcal{C}_{110}(t,t')$, and $\mathcal{C}_{010}(t)$ before computing the integrals. The second-order commutator can be written,
\begin{align*}
    \mathcal{C}_{10}(t) &= A(t)\frac{\Omega}{\sqrt{2}}\sum_{ij}z_{ij}^0\Big( \left[L_x, L_z^i\right]\sotimes L_z^j + L_z^i \sotimes \left[L_x , L_z^j\right]\Big)  \\
    &\equiv A(t)M_2.
\end{align*}
The third-order commutators are,
\begin{align*}
    \mathcal{C}_{010}(t) &= -A(t)\frac{\Omega_0}{\sqrt{2}}\sum_{ij}z_{ij}^0 \sum_{kl}z_{kl}^0\Big( \Big[L_z^{i},\left[L_x, L_z^k\right]\Big] \sotimes L_z^{l+j} \\
    &+ L_z^{k+i}\sotimes \Big[L_z^j,\left[L_x, L_z^l\right]\Big]\Big) \\
    &\equiv A(t)M_{3,1},
\end{align*}
\begin{widetext}
\begin{align*}
    \mathcal{C}_{110}(t,t') = A(t)A(t')\frac{\Omega_0}{\sqrt{2}}\sum_{ij}z_{ij}^0\Bigg\{&\frac{\Omega_0}{\sqrt{2}}\Big(\Big[L_x,\left[L_x, L_z^i\right]\Big]\sotimes L_z^j
    + L_z^i \sotimes \Big[L_x, \left[L_x , L_z^j\right]\Big]\Big)\\
    &+\Delta_0 \Big(\Big[L_z^2,\left[L_x, L_z^i\right]\Big]\sotimes L_z^j + L_z^i \sotimes \Big[L_z^2, \left[L_x , L_z^j\right]\Big]\Big)\\
    &+ \sum_{kl}\delta z_{kl} \Big(\Big[L_z^k,\left[L_x, L_z^i\right]\Big]\sotimes L_z^{j+l}+L_z^{i+k} \sotimes \Big[L_z^l, \left[L_x , L_z^j\right]\Big]\Big)\\
    & + 2 \frac{\Omega_0}{\sqrt{2}}\left[L_x, L_z^i\right]\sotimes \left[L_x, L_z^j\right]\Bigg\}  \\
    &\equiv A(t)A(t')M_{3,2}.
\end{align*}
\end{widetext}

With that in hand, the first-order term in the Magnus expansion can be written as
\begin{equation}
\begin{aligned}
    \bar{H}^{(1)} &= \frac{B(T)}{T}\Bigg[\left(\Delta_0 L_z^2 + \frac{\Omega_0}{\sqrt{2}}L_x\right)\sotimes I
    \\&+ I\sotimes \left(\Delta_0 L_z^2 + \frac{\Omega_0}{\sqrt{2}}L_x\right)
    + \sum_{ij} \delta z_{ij}L_z^i\sotimes L_z^j\Bigg] \\
    &+ \sum_{ij} z_{ij}^0 L_z^i\sotimes L_z^j\\
    &=M_1 + H_0,\label{eq:pulse_avg_hamiltonian1}
\end{aligned}
\end{equation}
where $B(t)\equiv\intf{0}{t}{A(t')}{t'}$ and $B(T)=T$ by definition. The second-order term becomes:
\begin{equation}
\begin{aligned}
   \bar{H}^{(2)} &= \frac{M_2}{2iT} \intf{0}{T}{\intf{0}{t}{A(t) -A(t')}{t'}}{t} \\
   &=\frac{M_2}{2iT}\intf{0}{T}{\left(tA(t) -B(t)\right)}{t}.
\end{aligned}
\end{equation}
For any $A(t)$ symmetric around $t=T/2$, this integral evaluates to $H^{(2)} = 0$.
And finally, the third-order term can be written as
\begin{equation}
\begin{aligned}
    \bar{H}^{(3)} &= \frac{-1}{6T}\intf{0}{T}{\intf{0}{t}{\intf{0}{t'}{\Big(\mathcal{C}_{110}(t,t')+\mathcal{C}_{110}(t'',t')\\
    &+\mathcal{C}_{010}(t)-2\mathcal{C}_{010}(t')+\mathcal{C}_{010}(t'')\Big)}{t''}}{t'}}{t} \\
    & = \frac{-1}{6T}\intf{0}{T}{\intf{0}{t}{\intf{0}{t'}{ \Big(A(t)A(t') + A(t')A(t'')\Big)M_{3,2}\\
    &+\Big(A(t)-2A(t') + A(t'')\Big)M_{3,1}}{t''}}{t'}}{t}\\
    &= \alpha(r, T)M_{3,2} + \beta(r, T)M_{3,1},
\end{aligned}
\end{equation}
where $\alpha(r, T)$ and $\beta(r, T)$ are functions of the ramp fraction and the evolution time only and, for the envelope defined in Eq.~\eqref{eq:analog_envelope}, are given by
\begin{equation*}
    \alpha(r,T) = T^2 \frac{(33-6\pi^2)r^3 + 3r^2(\pi^2-8) + 6\pi^2r - 4\pi^2}{72\pi^2(1-r)^2}
\end{equation*}
\begin{equation*}
    \beta(r,T) = \frac{rT^2}{12}\Bigg[1+\left(\frac{12}{\pi^2}-2\right)r\Bigg].
\end{equation*}

Taking the fact that all parameters in the Hamiltonian for the analog-digital two-qutrit experiment were on the order of $\qty{100}{\kHz}$, and using the evolution time $T=T_E=\qty{800}{\ns}$, we may estimate the total contribution of $\bar{H}^{(3)}$ to be $\left|\bar{H}^{(3)}\right|< \qty{0.1}{\kHz}$.

We now compute the errors due to higher-order terms in the dynamical decoupling sequence. We assume that the sequence alternates between time-independent Hamiltonians $\bar{H}_\mathrm{sys}$ and $\tilde{H}_\mathrm{sys}=U_{02}^\dagger \bar{H}_\mathrm{sys} U_{02}$, where $U_{02}=\pi_{02}\otimes \pi_{02}$ is considered to be instantaneous and perfect. Considering the order of magnitude of the error terms found above, we will simplify the calculation with the assumption that $\bar{H}_\mathrm{sys}$ is well approximated by the first order term Eq.~\eqref{eq:pulse_avg_hamiltonian1}. We can then use the discrete effective average Hamiltonian $\bar{H}_\mathrm{disc} \approx \bar{H}^{(1)}_\mathrm{disc}+\bar{H}^{(2)}_\mathrm{disc}$ with Hamiltonian terms defined in Eqs.~\eqref{eq:aht_discrete1}--\eqref{eq:aht_discrete2}.

The first order term,
\begin{equation}
\begin{aligned}
    \bar{H}^{(1)}_\mathrm{disc} &= \left(\Delta_0 L_z^2 + \frac{\Omega_0}{\sqrt{2}}L_x\right)\sotimes I + I\sotimes \left(\Delta_0 L_z^2 + \frac{\Omega_0}{\sqrt{2}}L_x\right)\\
    &+ \tilde{z}_{11}L_z\sotimes L_z + \tilde{z}_{22}L_z^2\sotimes L_z^2, \\
\end{aligned}
\end{equation}
produces the target Hamiltonian $H_\mathrm{AHM}$ when we account for the calibration $\tilde{z}_{22}\approx 0$. To compute the second order term, we first observe that the only terms affected by conjugating with $U_{02}$ are the odd-parity interaction terms $L_z L_z^2$ and $L_z^2 L_z$, and second, we observe that the only terms that do not commute with those are the on-site drives $L_x$. This simplifies the calculation of the commutator:
\begin{equation}
\begin{aligned}
    \bar{H}^{(2)}_\mathrm{disc} =& -\frac{\Omega_0 T}{2}\Big\{\tilde{z}_{12}\left(L_y\sotimes L_z^2 + L_z\sotimes [L_x,L_z^2]\right) \\
    &+ \tilde{z}_{21}\left([L_x,L_z^2]\sotimes L_z + L_z^2\sotimes L_y\right)\Big\}
\end{aligned}
\end{equation}
\revision{The order of magnitude of these errors, given the parameters of the two-qutrit experiment, is roughly 10\% of the $\tilde{z}_{11}$ interaction strength. Time-independent numerical simulations of the 2nd-order average Hamiltonian show that this error term produces an analog evolution unitary fidelity of $\sim\!97\%$ after a single Floquet step, relative to the ideal target Hamiltonian. This is the dominant coherent error term.}

\revision{In addition to coherent errors due to higher-order terms in the Magnus expansion, we have estimated the infidelity of the $\pi_{02}$ pulses used in this experiment to be $1-\mathcal{F}\sim 0.5\%$, based on repeated application of the gate. Numerical simulations of the gate on two qutrits suggest that this degree of infidelity is driven primarily by decoherence during the \qty{60}{\ns} $\pi_{02}$ pulses and ``quantum crosstalk'' arising from the static dispersive interaction \cite{sheldon_procedure_2016}. While these interaction terms are desired during the long periods of analog simulation, they are a source of error during the digital gates between those periods. We find minimal leakage to higher transmon states due to the relatively low pulse amplitudes and the use of DRAG pulses optimized for leakage prevention \cite{li_universal_2025, gambetta_analytic_2011}.}

\revision{Remaining error sources we have not analyzed rigorously include the breakdown of the instantaneous frame-change assumption due to non-zero duration of the $U_{02}$ pulses \cite{tyler_higher-order_2023, Zhou2024} and the validity of the adiabatic condition during the Stark drive ramps. We expect these error sources to be a relatively small contribution to the total error based on the good agreement to the time-independent average Hamiltonian simulations.}

\section{Device-specific details of the digital quantum simulator}\label{app_gate_approx}
In this appendix, we provide the matrix representation of the quantum operations that are the building blocks of our theory and provide the basis gate-set that are available in the AQT device where we performed our digital quantum simulation.

Single qutrit operations in the Abelian Higgs model are realized by the following quantum operations
\begin{equation}
    M_z =\mathrm{diag} (\exp(i\theta),1,\exp(i\theta) )
\end{equation}
and,
\begin{equation}
M_x ={\left(\begin{array}{ccc}
\cos^2 (\frac{\phi}{2\sqrt{2}} ) & \frac{-i}{\sqrt{2}} \sin (\frac{\phi}{\sqrt{2}} ) & -\sin^2 (\frac{\phi}{2\sqrt{2}} )\\
\frac{-i}{\sqrt{2}}\sin (\frac{\phi}{\sqrt{2}} ) & \cos (\frac{\phi}{\sqrt{2}}) & \frac{-i}{\sqrt{2}}\sin (\frac{\phi}{\sqrt{2}}) \\
-\sin^2 (\frac{\phi}{2\sqrt{2}} ) & \frac{-i}{\sqrt{2}} \sin (\frac{\phi}{\sqrt{2}} ) & \cos^2 (\frac{\phi}{2\sqrt{2}} )
\end{array}\right)} ;
\end{equation}
whereas nearest-neighbor spin-1 Ising interaction terms are the entangling operations
\begin{equation}
  M_{zz}=\mathrm{diag}\left(
e^{i \gamma},\ 1, \ e^{-i \gamma},\ 1, 1,\ 1, e^{-i \gamma} ,1, e^{i \gamma} \right) .\label{eqn_EEop}
\end{equation}
We discussed the unitary decomposition of the above quantum operations into elementary gates in the main text. For the decomposition, we used the Gell-Mann matrices which are a standard choice of generators for $\mathrm{SU}(3)$:
\begin{widetext}
\small{\begin{align*}
& \lambda_1 \equiv \sigma_x^{01}=\left(\begin{array}{ccc}
0 & 1 & 0 \\
1 & 0 & 0 \\
0 & 0 & 0
\end{array}\right), \quad \lambda_2 \equiv \sigma_y^{01}=\left(\begin{array}{ccc}
0 & -i & 0 \\
i & 0 & 0 \\
0 & 0 & 0
\end{array}\right), \quad \lambda_3 \equiv \sigma_z^{01}=\left(\begin{array}{ccc}
1 & 0 & 0 \\
0 & -1 & 0 \\
0 & 0 & 0
\end{array}\right) \text {, } \\
& \lambda_4 \equiv \sigma_x^{02}=\left(\begin{array}{ccc}
0 & 0 & 1 \\
0 & 0 & 0 \\
1 & 0 & 0
\end{array}\right), \quad
\lambda_5 \equiv \sigma_y^{02}=\left(\begin{array}{ccc}
0 & 0 & -i \\
0 & 0 & 0 \\
i & 0 & 0
\end{array}\right), \quad \lambda_6 \equiv \sigma_x^{12}=\left(\begin{array}{ccc}
0 & 0 & 0 \\
0 & 0 & 1 \\
0 & 1 & 0
\end{array}\right) \text {, } \\
& \lambda_7 \equiv \sigma_y^{12}=\left(\begin{array}{ccc}
0 & 0 & 0 \\
0 & 0 & -i \\
0 & i & 0
\end{array}\right), \quad \lambda_8=\frac{1}{\sqrt{3}}\left(\begin{array}{ccc}
1 & 0 & 0 \\
0 & 1 & 0 \\
0 & 0 & -2
\end{array}\right) . \\
&
\end{align*} }
\end{widetext}
These are Hermitian matrices with zero trace, and orthogonal under trace.  The Gell-Mann matrices $(\lambda_i)$ constitute a viable set of generators of $\mathrm{SU}(3)$, providing universal control for a single qutrit. For convenience, we define an additional Hermitian operator, $\sigma_z^{12}\equiv-\frac{1}{2}\lambda_3-\frac{\sqrt{3}}{2}\lambda_8$, which is not orthogonal to $\sigma_z^{01}$, but which corresponds to qubit-like rotations about the $z$ axis in the $\{\ket{1},\ket{2}\}$ subspace and is naturally implemented by shifting the phase of the drive on the $\ket{1}\leftrightarrow\ket{2}$ transition.

We can use either the two-qudit controlled-\textsc{sum} or controlled-Z gates for realizing the entangling operations, given by \cite{gottesman_fault-tolerant_1999}
\begin{align}
U_\mathrm{CSUM} &=\sum_{n=0}^{d-1}|n\rangle\langle n| \sotimes X_d^n\\
&=\left(I \sotimes H_d^\dagger \right) U_\mathrm{CZ}\left(I \sotimes H_d \right) \\
U_\mathrm{CZ}  &=\sum_{n=0}^{d-1}|n\rangle\langle n| \sotimes Z_d^n,
\end{align}
where,
\begin{align}
X_d&=\sum_{n=0}^{d-1}\ket{{n+1}\imod{d}}\bra{n}, \\
Z_d&=\sum_{n=0}^{d-1} \exp\Big(i\frac{2\pi n}{d} \Big) \ketbra{n}{n}, \\
H_d&=\frac{1}{\sqrt{d}} \sum_{n,m=0}^{d-1} \exp\Big(i\frac{2\pi}{d}nm \Big) \ketbra{n}{m} .
\end{align}
The $\mathrm{CSUM}$ and $\mathrm{CZ}$ gates are the qudit generalization of the qubit CNOT and controlled-Z gate.

Our native gate for two-qutrit operations is the qutrit-CZ gate defined by the unitary
\begin{equation}
    U_\mathrm{CZ} = \sum_{i,j=0}^2\omega^{ij}\ketbra{ij}{ij}
\end{equation}
where $\omega = e^{i2\pi/3}$ is the third root of unity. To realize the gate on our device, we employ simultaneous off-resonant microwave drives to realize a tunable cross-Kerr interaction in our coupled qutrit system and employ dynamical decoupling to engineer the specific target unitary of $U_\mathrm{CZ}$. More information on how the gate is performed can be found in Ref.~\cite{goss_high-fidelity_2022}. We use the same principle without the requirement that the final unitary is $U_\mathrm{CZ}$ to engineer the interaction Hamiltonian for the analog-digital hybrid simulation.

\begin{table}[h!]
\centering
\begin{tabular}{l|ccc}
\hline
 & QA && QB \\
\hline
Qubit freq. (GHz) & 5.374 && 5.443 \\
Anharm. (MHz) & -274 && -271 \\
$T_1^{01}~(\mu\text{s})$ & 60(13) && 47(14) \\
$T_1^{12}~(\mu\text{s})$ & 36(4) && 34(5) \\
$T_{2e}^{01}~(\mu\text{s})$ & 69(7) && 61(14) \\
$T_{2e}^{12}~(\mu\text{s})$ & 32(5) && 32(6) \\
$\tau_{1Q,X90}$ (ns) & 30 && 30 \\
$\tau_{1Q,\mathrm{SU}(3)}$ (ns) & 180 && 180 \\
$\tau_{2Q}$ (ns) && 612 &  \\
\hline
\end{tabular}
\caption{Device parameters for the Berkeley based qutrit processor used in the digital quantum simulations. Decoherence times ($T$), various native and composite gate times ($\tau$) are listed along with qubit frequencies and anharmonicities used in the digital experiments. \label{table_Berkeley}}
\end{table}

\begin{figure*}[!htb]
    \begin{centering}
    \subfloat{\includegraphics[width=0.45\linewidth]{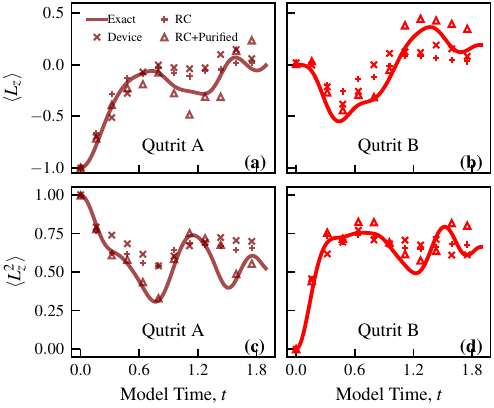} }
    \subfloat{\includegraphics[width=0.45\linewidth]{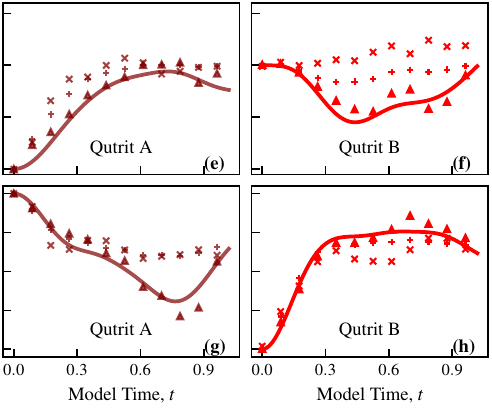} }
    \end{centering}
    \caption{ Impact of error mitigation scheme is investigated in two experiments with different Trotter steps. A larger Trotter step ($2\pi\,\delta t =1.0$) allows us to observe  real-time dynamics longer times (a-d), whereas a smaller Trotter step ($2\pi\,\delta t =0.55$) reduces Trotter error but limits the accessible timescale (e-h). In both experiments, we find error mitigation schemes to be effective in improving fidelity of our quantum operations on average. For both experiments, we present here raw hardware data (Device), error-mitigated device data with randomized compilation (RC), data with randomized compilation and numerical purification (RC+purified), and compare them with exact diagonalization (Exact) for qutrit A (Maroon) and qutrit B (Red). Expectation values are computed for an initial state $\ket{21}$. We note that statistical errors in our experiments are insignificant compared to other sources of error in the experiment.  Statistical error scales as the inverse of the square root of the number of shots and thus statistical errors in our experiment are of the order $\sim \frac{1}{\sqrt{1024\times 30}} \sim 0.006$.}
    \label{fig_app_2Q_lz_lzsqr}
\end{figure*}

\section{Error mitigation in qutrit-based simulation}\label{app_digital_mitigation}

\begin{figure*}[!htb]
\includegraphics[width=1\linewidth]{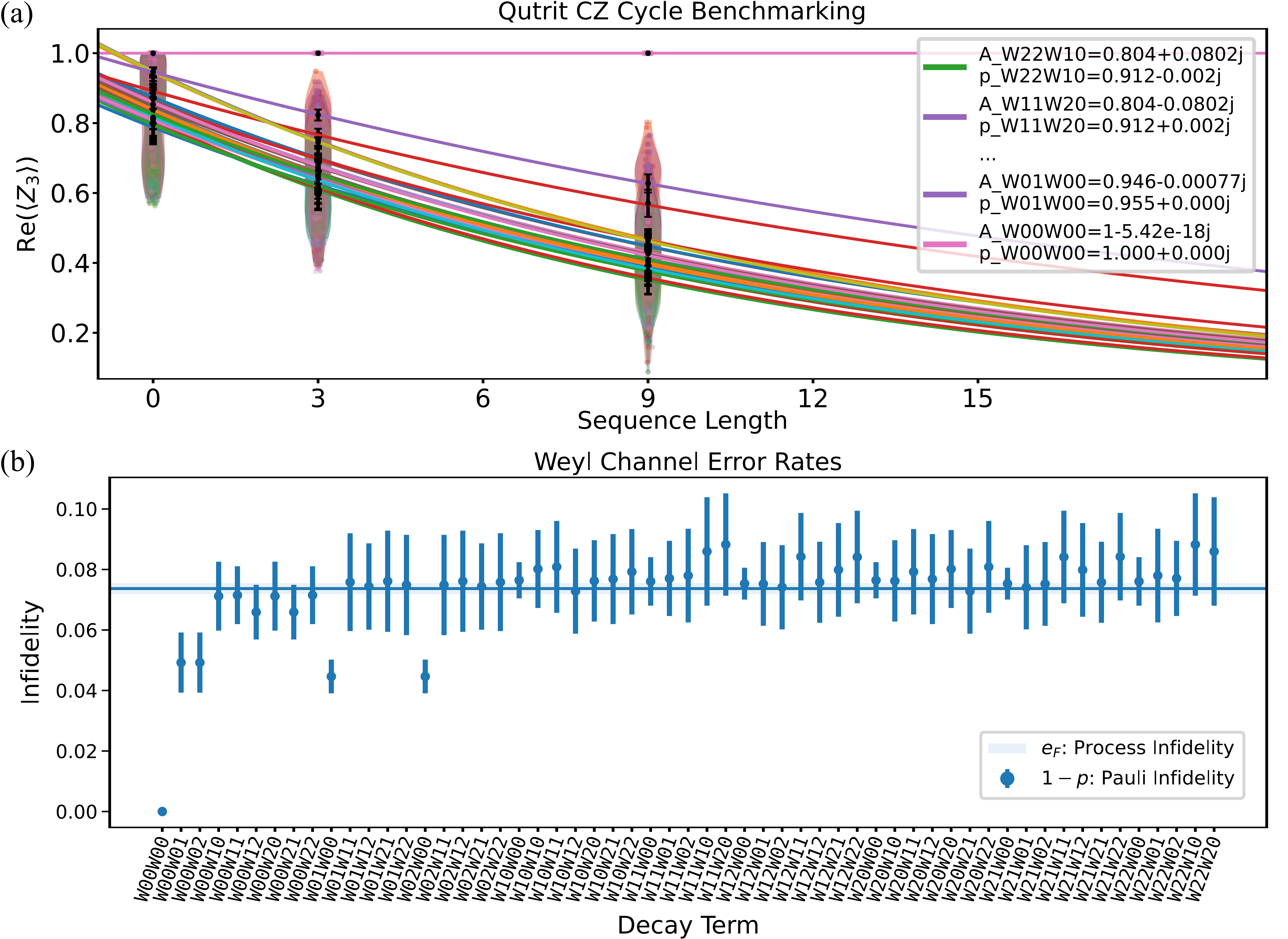}
\caption{Cycle Benchmarking results for our two-qutrit CZ gate employed as the native entangling gate in the digital quantum simulations of the Abelian Higgs model. (a) The violin plots of individual Weyl decay exponentials and their associated uncertainty. Both the SPAM parameter (A) and the decay term (p) are fit for each prepared Weyl eigenstate. (b) The infidelity and uncertainty associated with each decay term extracted from fitting the Weyl eigenstate decays from our CB circuits. Cycle benchmarking measures the combined error of the CZ gate and the random Weyl gates used for twirling.}
\label{fig:CB_fig}
\end{figure*}
In this appendix section, we review the error mitigation scheme developed in earlier studies \cite{Goss:2023frd, PhysRevA.108.032417} which proved instrumental for improving the time scale over which we could observe dynamics. Our error strategy is two-fold: firstly we apply randomized compilation to tailor our error model to be stochastic Pauli noise (approximately depolarizing), and secondly we numerically purify our expectation values using noise measured by cycle benchmarking. In addition, readout error mitigation is applied to reduce state preparation and measurement errors. Data from the digital experiment is shown in Fig.~\ref{fig_app_2Q_lz_lzsqr} both with and without mitigation steps to demonstrate the improvement of the measured populations.\\

We can construct a stochastic channel efficiently by sampling from a unitary-1 design \cite{Graydon:2022nfb}.
We choose Weyl (qudit Pauli) operators as the unitary-1 design to perform twirling for randomized compilation. Weyl operators form a group over $\mathrm{SU}(d)$ with group elements defined as $W_{k,l}= \omega^{-kl/2} Z^k X^l $ where X and Z gates are defined by their action on the occupation basis as $\ket{m}=\{\ket{0},\ket{1},\ket{2}\}$ such that $X\ket{m}=\ket{m\oplus 1}$ and $Z\ket{m}=\omega^m \ket{m}$, with $\omega=\exp(\frac{i2\pi}{3})$. It is worth mentioning that using higher dimensional qudits as our quantum computing unit does not increase the overhead (number of twirling circuits) required to achieve equal suppression of coherent noise with randomized compilation \cite{Goss:2023frd}. To achieve effective coherent error suppression, we perform 30 equivalent twirled circuits for each initial circuit. In principle, employing randomized compiling does not incur a runtime overhead as the number of desired shots per circuit can be divided evenly over the twirled circuits.

Our strategy after performing randomized compiling to mitigate coherent noise is to learn and numerically purify the incoherent noise in our system. Formally, incoherent noise can be defined in terms of error rates associated with Weyl channels. A Weyl channel can be expressed as
\begin{align}
    \mathcal{E}(\rho) &= \sum_{p,q=0}^{2} w_{p,q} W_{p,q} \rho W_{p,q}^\dagger \nonumber \\
    &= w_{0,0} \rho + \sum_{p,q,p+q\neq 0} w_{p,q} W_{p,q} \rho W_{p,q}^\dagger
\end{align}
We efficiently learn the error rate $w_{p,q}$ of the Weyl channels $W_{p,q}$ using cycle benchmarking \cite{Erhard:2019cxk}. Cycle benchmarking is performed by executing the following steps
\begin{itemize}
   \item Prepare eigenstates of a randomly chosen Weyl channel.
   \item Select a cycle (e.g. our native two-qudit gate) and repeat it $N$-times. In between each cycle perform Weyl twirling to turn the channel into a stochastic one.
   \item Measure the expectation value of the same Weyl operator after $N$ times.
\end{itemize}
Repeating the steps for several layers ($N$) and initial choice of Weyl eigenstates, the decay rate of each associated channel can be measured by fitting its corresponding error curve. The cycle benchmarking results of our two-qutrit CZ gate can be found in Fig. ~\ref{fig:CB_fig}.

We can estimate an effective depolarizing noise parameter from the average of our Weyl decay terms measured by cycle benchmarking. We note this simplified depolarizing noise model for our system takes the form,
\begin{equation}
    \epsilon(\rho) = \lambda \rho + (1-\lambda)\frac{I_3}{3^N}
\end{equation}
As can be observed in Fig.~\ref{fig:CB_fig}, the fairly even distribution of Weyl decay terms makes this a reasonable approximation. To mitigate our incoherent noise, the estimated depolarization strength is used to rescale the $Z$ expectation values measured from our twirled circuits, providing a more accurate estimate of the observables. For a detailed discussion on Weyl twirling and cycle benchmarking, readers are advised to consult \cite{Goss:2023frd, Hashim:2024dox}.

The impact of applying these mitigation schemes proved instrumental in recovering useful dynamics. Fig~\ref{fig_app_2Q_lz_lzsqr} demonstrates how randomized compilation and purification improved the estimate of the field operator estimates $\langle L_z \rangle$ and $\langle L_z^2 \rangle$. It is clear that twirling itself is not always sufficient to recover low-error estimates as decoherence dominates beyond very shallow circuit depths. A combination of mitigation schemes, namely randomized compilation and purification, proved essential in reducing coherent and incoherent errors.\\

\section{Resource estimation and comparison with qubit-based simulation}\label{app_qubit_resource_estimate}
We provide arguments in this appendix on the advantages of qutrit-based time-evolution of spin-1 quantum system over qubit-based formulations.

To map an $N$-qutrit three level system in a qubit platform, we require $ N \log_{2} 3$ qubits. A possible mapping of the spin-1 eigenstates to the occupation basis using standard binary encoding is
\begin{align}
    &\ket{m=1} \to \ket{n=00}, \quad
    \ket{m=0} \to \ket{n=01}, \nonumber \\
    &\ket{m=-1}\to \ket{n=10}.
\end{align}
With this encoding, different quantum operations required to implement the propagator of the AHM are shown in Fig.~\ref{fig_qubit_AHM}. We did not investigate other encoding schemes for this study. It is worth mentioning that there is no unique encoding scheme that is advantageous for all sorts of different quantum operations \cite{Sawaya:2020bky}.
We performed both analytic and transpiler-based decomposition for each quantum operation for the standard binary encoding. We find the decomposition obtained from the Bqskit transpiler \cite{doecode_58510} optimal with fewer entangling operations for all the cases. Two-qubit unitary synthesis of our single qutrit operation $M_x^i$ requires two CNOT gates, whereas a 4-qubit decomposition of the original entangling operation $M_{zz}^i$ requires 12-CNOT gates if we consider all-to-all topology of qubits in the hardware. For a restricted topology, for example with IBM's Heron r3 processor, we require 30 CNOT gates. This clearly demonstrates that for a mature technology with comparable decoherence times, qutrit simulation would be advantageous to that of qubit based simulation of the Abelian Higgs and other $\mathrm{SU}(3)$ based gauge theories, and various spin-1 models.

\begin{turnpage}
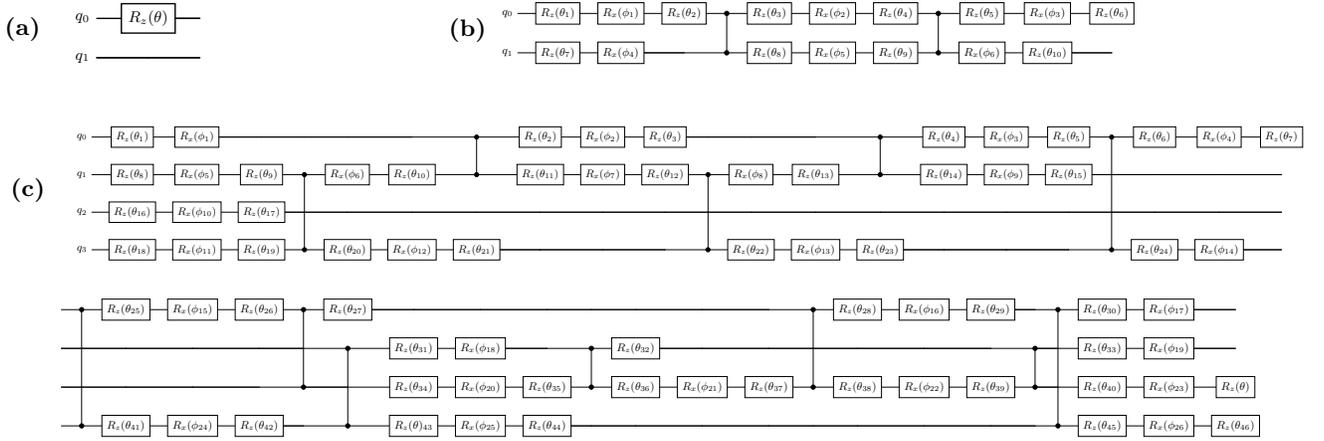
\begin{figure*}[p]
\centering
\begin{minipage}{0.18\linewidth}
\centering
\textbf{(a)}\quad
\resizebox{0.6\linewidth}{!}{
\begin{quantikz}
\lstick{$q_0$} & \gate{R_z(\theta)} & \qw \\
\lstick{$q_1$} & \qw & \qw
\end{quantikz}
}
\end{minipage}
\hfill
\begin{minipage}{0.8\linewidth}
\centering
\textbf{(b)}
\resizebox{0.8\linewidth}{!}{
\begin{quantikz}
\lstick{$q_0$} & \gate{R_z(\theta_1)} & \gate{R_x(\phi_1)} & \gate{R_z(\theta_2)} & \ctrl{1} & \gate{R_z(\theta_3)} & \gate{R_x(\phi_2)} & \gate{R_z(\theta_4)}  & \ctrl{1} & \gate{R_z(\theta_5)} & \gate{R_x(\phi_3)} & \gate{R_z(\theta_6)}  \\
\lstick{$q_1$} & \gate{R_z(\theta_7)} & \gate{R_x(\phi_4)} & \qw & \control{} & \gate{R_z(\theta_8)} & \gate{R_x(\phi_5)} & \gate{R_z(\theta_9)} & \control{} & \gate{R_x(\phi_6)} & \gate{R_z(\theta_{10})} & \qw
\end{quantikz}
}
\end{minipage}

\vspace{0.7cm}

\begin{minipage}{\linewidth}
\centering
\textbf{(c)} \vspace{\baselineskip}

\resizebox{0.99\linewidth}{!}{
\begin{quantikz}
\lstick{$q_0$} & \gate{R_z(\theta_1)} & \gate{R_x(\phi_1)} & \qw        &\qw &\qw  &\qw & \ctrl{1} & \gate{R_z(\theta_2)} & \gate{R_x(\phi_2)} & \gate{R_z(\theta_3)} & \qw &\qw &\qw & \ctrl{1} & \gate{R_z(\theta_4)} & \gate{R_x(\phi_3)} & \gate{R_z(\theta_5)} & \ctrl{3} & \gate{R_z(\theta_6)} & \gate{R_x(\phi_4)} & \gate{R_z(\theta_7)} \\
\lstick{$q_1$} & \gate{R_z(\theta_8)} & \gate{R_x(\phi_5)} & \gate{R_z(\theta_{9})} &\ctrl{2} &\gate{R_x(\phi_6)} &\gate{R_z(\theta_{10} )} & \control{} & \gate{R_z(\theta_{11} )} & \gate{R_x(\phi_7)} & \gate{R_z(\theta_{12} )} & \ctrl{2} &\gate{R_x(\phi_8)} &\gate{R_z(\theta_{13} )} & \control{} & \gate{R_z(\theta_{14} )} & \gate{R_x(\phi_9)} & \gate{R_z(\theta_{15} )} & \qw & \qw & \qw & \qw \\
\lstick{$q_2$} & \gate{R_z(\theta_{16} )} & \gate{R_x(\phi_{10} )} & \gate{R_z(\theta_{17} )} &\qw      &\qw        &\qw         & \qw        & \qw        & \qw       & \qw         & \qw      &\qw        &\qw        & \qw       & \qw        & \qw            & \qw        & \qw        & \qw        & \qw     & \qw \\
\lstick{$q_3$} & \gate{R_z(\theta_{18} )} & \gate{R_x(\phi_{11} )} & \gate{R_z(\theta_{19} )}  &\control{} &\gate{R_z(\theta_{20} )} &\gate{R_x(\phi_{12})} & \gate{R_z(\theta_{21} )}  & \qw & \qw & \qw &\control{}    &\gate{R_z(\theta_{22} )} & \gate{R_x(\phi_{13} )} & \gate{R_z(\theta_{23} )} & \qw & \qw & \qw &\control{} & \gate{R_z(\theta_{24} )} & \gate{R_x(\phi_{14} )} & \qw
\end{quantikz}
}
\end{minipage}

\vspace{0.35cm}

\begin{minipage}{0.9\linewidth}
\centering
\hspace{5pt}
\resizebox{\linewidth}{!}{
\begin{quantikz}
 \qw & \ctrl{3} &\gate{R_z(\theta_{25})} &\gate{R_x(\phi_{15})} &\gate{R_z(\theta_{26})} &\ctrl{2} &\gate{R_z(\theta_{27})} & \qw & \qw & \qw & \qw & \qw & \qw & \qw &\ctrl{2} &\gate{R_z(\theta_{28})} &\gate{R_x(\phi_{16})} &\gate{R_z(\theta_{29})} & \qw & \ctrl{3}  & \gate{R_z(\theta_{30})} &\gate{R_x(\phi_{17} )} &\qw \\
 \qw & \qw &\qw &\qw &\qw &\qw &\ctrl{2} & \gate{R_z(\theta_{31})} & \gate{R_x(\phi_{18} )} & \qw   & \ctrl{1} & \gate{R_z(\theta_{32} )} & \qw & \qw &\qw &\qw &\qw &\qw & \ctrl{1} & \qw  & \gate{R_z(\theta_{33} )} &\gate{R_x(\phi_{19} )} &\qw \\
 \qw & \qw &\qw &\qw &\qw &\control{} &\qw & \gate{R_z(\theta_{34} )} & \gate{R_x(\phi_{20} )} & \gate{R_z(\theta_{35})}   & \control{} & \gate{R_z(\theta_{36})} & \gate{R_x(\phi_{21} )} & \gate{R_z(\theta_{37})} &\control{} &\gate{R_z(\theta_{38} )} &\gate{R_x(\phi_{22} )} &\gate{R_z(\theta_{39} )} & \control{} & \qw  & \gate{R_z(\theta_{40} )} &\gate{R_x(\phi_{23} )} &\gate{R_z(\theta)} \\
 \qw & \control{} &\gate{R_z(\theta_{41} )} &\gate{R_x(\phi_{24})} &\gate{R_z(\theta_{42} )} &\qw &\control{} & \gate{R_z(\theta_{43}) } & \gate{R_x(\phi_{25})} & \gate{R_z(\theta_{44} )} & \qw & \qw & \qw & \qw &\qw &\qw &\qw &\qw & \qw & \control{}  & \gate{R_z(\theta_{45} )} &\gate{R_x(\phi_{26})} &\gate{R_z(\theta_{46})}
 \end{quantikz}
}
\end{minipage}
\caption{Transpiler-compiled qubit-gate decomposition of the different operations in the Trotterized evolution of the spin-1 truncated AHM. With binary encoding, decomposition of the (a) $M_z$, (b) $M_x$, and (c) $M_{zz}$ operations are shown. Numerical rotation angles of different operations are replaced with symbols for visual clarity. Note the increase in the number of entangling operations for the qubit decomposition compared to the qutrit decomposition and the presence of non-local entangling operations in the qubit formulation. In the absence of all-to-all connectivity of the qubits, the number of entangling operation increases further due to the presence of additional swap gates between qubit pairs. Each SWAP operation introduces three entangling gates, thus rendering the qubit decomposition even more costly.}
    \label{fig_qubit_AHM}
\end{figure*}
\end{turnpage}

\begin{figure*}[!htb]
\captionsetup[subfigure]{labelformat=nocaption}
\includegraphics[width=0.9\linewidth]{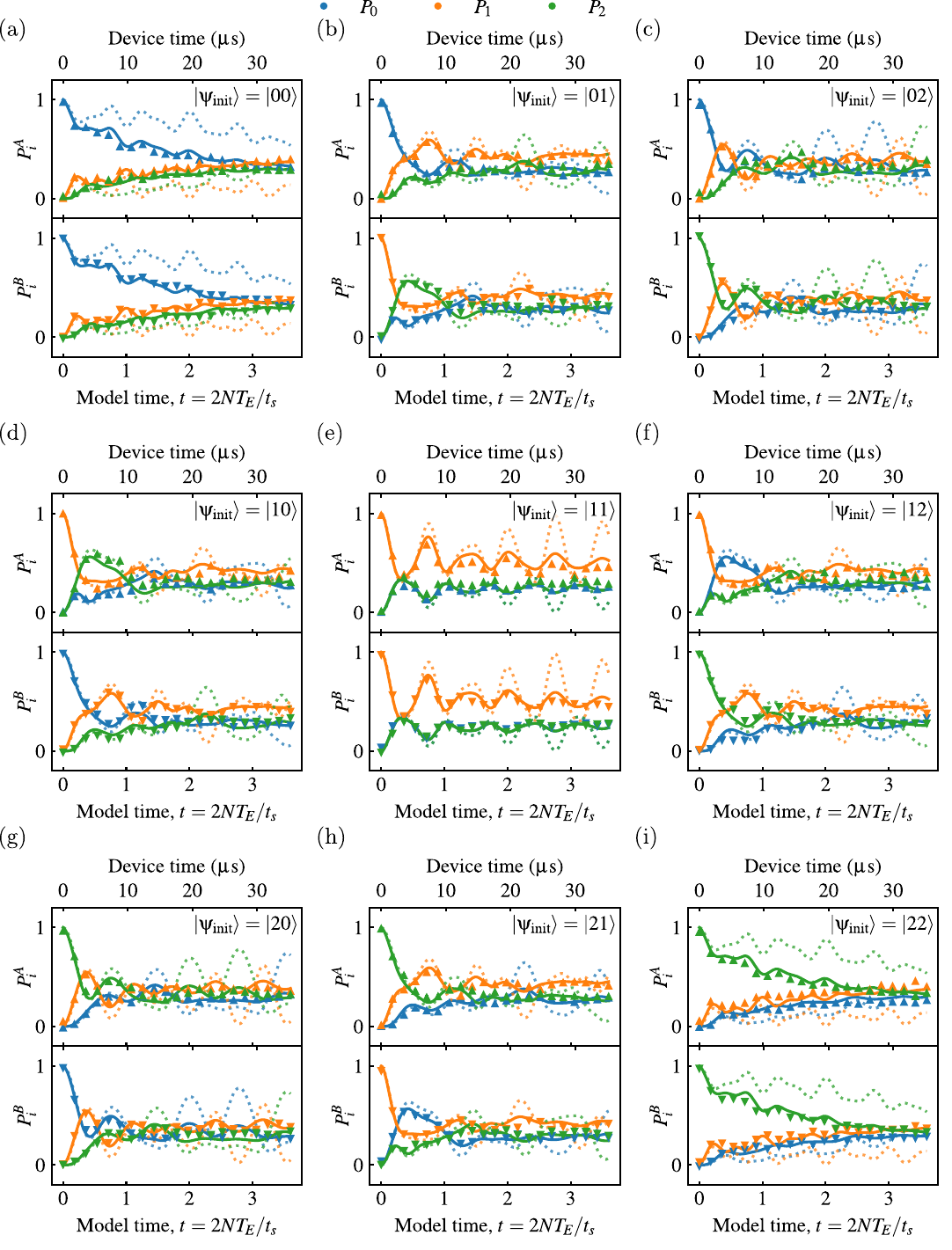}%
\begin{subfigure}{0\linewidth}
\caption{}\label{fig:analog_data_00}
\end{subfigure}%
\begin{subfigure}{0\linewidth}
\caption{}\label{fig:analog_data_01}
\end{subfigure}%
\begin{subfigure}{0\linewidth}
\caption{}\label{fig:analog_data_02}
\end{subfigure}%
\begin{subfigure}{0\linewidth}
\caption{}\label{fig:analog_data_10}
\end{subfigure}%
\begin{subfigure}{0\linewidth}
\caption{}\label{fig:analog_data_11}
\end{subfigure}%
\begin{subfigure}{0\linewidth}
\caption{}\label{fig:analog_data_12}
\end{subfigure}%
\begin{subfigure}{0\linewidth}
\caption{}\label{fig:analog_data_20}
\end{subfigure}%
\begin{subfigure}{0\linewidth}
\caption{}\label{fig:analog_data_21}
\end{subfigure}%
\begin{subfigure}{0\linewidth}
\caption{}\label{fig:analog_data_22}
\end{subfigure}%
\caption{Two-qutrit analog-digital experimental results, extended. Data shown in Fig.~\ref{fig:analog_data_double} is excerpted from \subref{fig:analog_data_21}. Dotted lines show exact evolution of the AHM under the Scr\"odinger equation, solid lines show evolution under the Lindblad model described in Appendix~\ref{app:analog_error_terms}, and triangle markers are experimentally measured transmon eigenstate probabilities, after a statistical readout correction has been applied. Model parameters were $\kappa/2\pi=\chi/2\pi=\beta/2\pi=1$, and the scale frequency for all measurements was $t_s^{-1}\approx \qty{110}{\kHz}.$ Uncertainties are estimated from shot noise $\sigma\sim 1/\sqrt{N_\mathrm{shots}}$ where $N_\mathrm{shots}=3000$. Error bars are smaller than the markers.
}
\label{fig:analog_double_extended}
\end{figure*}

\clearpage

\clearpage
\bibliography{abelian_higgs}%

\end{document}